\documentclass[preprint,aps]{revtex4}
\usepackage[latin1]{inputenc}
\usepackage[dvips]{color}
\usepackage{graphicx, here}
\usepackage{amsfonts,amstext,amsbsy,amssymb,amsmath}

\def\be{\begin{equation}}
\def\ee{\end{equation}}

\newcommand{\weglassen}[1]{}

\newcommand{\const}{\text{const}}

\newcommand{\halb}{{\textstyle \frac{1}{2}}}

\renewcommand{\d}{\text{d}}

\begin{document}

\title{Extended phase diagram of the Lorenz model}
\author{Holger R. Dullin$^{1,2}$, Sven Schmidt$^2$, Peter H. Richter$^2$, Siegfried K. Grossmann$^3$}
\affiliation{%
$^1$Mathematical Sciences Loughborough University, LE11 3TU, Loughborough, UK\\
$^2$Institut fuer Theoretische Physik der Universit\"at Bremen, D-28334 Bremen, Germany\\ 
$^3$Fachbereich Physik der Philipps-Universit\"at, Renthof 6, D-35032 Marburg, Germany\\}

\date{\today}

\begin{abstract}
The parameter dependence of the various attractive solutions of the three
variable nonlinear Lorenz model equations for thermal convection in Rayleigh-B\'enard
flow is studied. Its bifurcation structure has commonly been investigated
as a function of~$r$, the normalized Rayleigh number, at fixed Prandtl
number~$\sigma$. The present work extends the analysis to the entire 
$(r,\sigma)$ parameter plane.
An onion like periodic pattern is found which is due to the alternating stability of
symmetric and non-symmetric periodic orbits. This periodic pattern is explained
by considering non-trivial limits of large~$r$ and~$\sigma$. In addition to
the limit which was previously analyzed by Sparrow, we identify two more distinct
asymptotic regimes in which either $\sigma / r$ or $\sigma^2 / r$ is
constant. In both limits the dynamics is approximately described by Airy functions
whence the periodicity in parameter space can be calculated analytically.
Furthermore, some observations about sequences of bifurcations and coexistence
of attractors, periodic as well as chaotic, are reported.

\end{abstract}


\keywords{Lorenz model, periodic orbits, bifurcations, large parameter asymptotics}

\maketitle

\vspace{8mm}

\section{Introduction}

The nonlinear coupled set of three ODEs
\be
  \label{lorenz}
    \dot x = -\sigma x + \sigma y  \  , \qquad
    \dot y = r x - y -xz  \ , \qquad
    \dot z = -bz + xy  \ ,
\ee
was first derived by
Berry Saltzman and Edward Lorenz \cite{Sa62,Lo63} as a model for
thermal convection between two plates perpendicular to the direction
of the earth's gravitational force. It has three control parameters.
First, there is the normalized temperature difference $r=\Delta T /
\Delta T_c = Ra / Ra_c$ between the hotter lower plate and the
colder upper one. $\Delta T_c$~is the temperature difference at the
onset of convection. $Ra = g \alpha_p L^3 \Delta T / \nu \kappa$,
the Rayleigh number, is the dimensionless temperature difference.
$g$~denotes the gravitational acceleration, $\alpha_p$~the isobaric
thermal expansion coefficient, $L$~the distance between the plates
in upward direction, and $\nu$, $\kappa$ characterize the kinematic
viscosity and thermal diffusivity, respectively.

Second, there is the Prandtl number $\sigma = \nu / \kappa$,
describing the ratio of the two molecular time scales
$\tau_{\nu} = L^2 / \nu$ and $\tau_{\kappa} = L^2 / \kappa$
for momentum and energy transport in the system.
$\sigma$~also measures the ratio of the widths $\lambda_{u}$ and $\lambda_{T}$
of the kinematic and the thermal boundary layers. Large Prandtl
numbers~$\sigma$ indicate a broad kinematic boundary layer containing a
small thermal one, fast molecular momentum exchange and small convective
Reynolds numbers. Small Prandtl numbers characterize fast molecular thermal
exchange, a broad thermal boundary layer containing a thinner kinematic one and
larger convective Reynolds numbers. Typical values for~$\sigma$ are
$0.7$ in air (at room temperature), N$_2$, O$_2$, He, and other gases,
about~7 in water (room temperature), $600$ through $3000$ in typical
organic liquids as e.g. dipropylene glycol~($600$), or
$\sigma \approx 7250$ in glycerine. Small Prandtl numbers are realized
in mercury ($\sigma \approx 0.025$) or liquid sodium with $\sigma
\approx 0.005$.

The third parameter in eqs.~(\ref{lorenz}) is the coupling strength~$b$,
originating from the nonlinearity $(\vec{u} \cdot \vec{\nabla}) T$ of
the Boussinesq equation. It is of order $1$ and conventionally
chosen as $b = 8/3$; we adopt this value here.

The variables $x(t), y(t)$, and $z(t)$ describe, respectively, the amplitude
of the velocity mode (the only one left in the model), the amplitude of the
corresponding temperature mode, and the amplitude of the mode measuring the
nondimensional heat flux~$Nu$, the Nusselt number. $x, y$
characterize the roll pattern, the heat flux is
\be
Nu = 1 + 2 z / r \ .
\label{nusselt}
\ee

For $0 \le r \le 1$ there exists one stable fixed point~P$_0$ ($x_0 = y_0 =
z_0 = 0$), with $Nu = 1$, corresponding to pure molecular heat conduction. For $r$
exceeding~$1$, convective rolls develop, described by two stable
fixed points
\be
 \label{fixedpoints}
  \text{P}_{\pm} = (x_{\pm},y_{\pm},z_\pm) = \Bigl( \pm \sqrt{b(r -1)}, \pm \sqrt{b(r -1)}, r-1 \Bigr)
\ee
hence $Nu = 3 - 2/r$. The trajectories in phase
space $\Gamma = \{x, y, z\}$ are spirals towards these fixed points.
Stability is lost for
\be
r > r_c = \sigma \;\frac{\sigma + b + 3}{\sigma -b -1}  \ .
\label{stabiloss}
\ee
For the canonical choice $\sigma = 10$ it is $r_c = 24{.}7368...$.
At such large~$r$ the three mode approximation for the PDEs describing thermal convection has of
course ceased to be physically realistic, but mathematically the model now starts to show
its most fascinating properties, because a strange
attractor develops for $r > r_c$. With further increasing~$r$, windows of regular behavior emerge out of chaos
where stable periodic orbits exist. These stability windows tend to
become broader and broader until for sufficiently large~$r$ chaos seems to disappear
and one type of stable periodic solution takes over.

Of the numerous studies of the Lorenz model we cite as examples the monograph~\cite{Sp82}
and the more recent survey in Jackson's book~\cite{Ja91}. To the best of our knowledge,
the parameter dependence of the Lorenz model has mainly been analyzed with respect to
its control parameter~$r$, the external forcing, and sometimes with varying~$b$ (as in
the work~\cite{AF84} which is cited in~\cite{Ja91}). The Prandtl number~$\sigma$ has
usually been fixed to its canonical value $\sigma = 10$, and as a rule, $b$ has been
taken as $8/3$. The limit of large~$r$ at arbitrary fixed~$\sigma$ was thoroughly studied by
Robbins~\cite{Ro79} and Sparrow~\cite{Sp82}.

In the present paper we study the parameter range where both~$r$ {\em and} $\sigma$ become large.
The scheme of bifurcations is then a very different one. With $\sigma$ as large as 60
(Fig.~\ref{fig:r409sigma60}) one is just beginning to see a structure that unfolds on
much larger scales (Fig.~\ref{fig:rsigma2000}). There is ongoing repetition of
a few basic features, but these are different from the structure at low~$\sigma$. The
line $\sigma_s = \sqrt{br}$ divides these two regimes as we shall show. Our theoretical explanation of these
findings focuses on a backbone structure of the dominant stable periodic attractors, but
orbits with arbitrarily high periods, and chaos, are also observed as part of the pattern.

An important feature of the Lorenz equations is their invariance with respect to the reflection
\be
     \Sigma :  \quad  x, y  \mapsto -x, -y  \quad \mbox{and} \quad  z \mapsto z  \ .
\label{symmetry1}
\ee
Evidently the $z$-axis is a $\Sigma$-invariant line. For $r>1$ it is also the stable manifold of the
fixed point~P$_0$. This characteristic symmetry property with respect to~$\Sigma$ has implications on the nature of
the period doubling scenarios which occur over and over again as the parameters $r$, $\sigma$ are varied:
they do \emph{not} follow the familiar scheme of the
paradigmatic logistic iteration $x \mapsto ax(1-x)$ but
rather the somewhat more intricate scheme of the cubic iteration $x\mapsto f(x)=(1-c)x + cx^3$.
This cubic mapping possesses the symmetry $f=\tilde\Sigma^{-1}f\tilde\Sigma$, with $\tilde\Sigma:\ x\mapsto -x$,
and may be viewed as one building block for understanding the transition to chaos in the Lorenz system,
see Hao et al.~\cite{Di90,Fa96} for a detailed analysis of this aspect.

The motivation of the present work was the simple question why
the meteorologist Lorenz would have chosen the Prandtl number $\sigma = 10$,
which is adequate for water but less so for air with its order of
magnitude $\sigma \approx 1$. The reader will easily find
that taking~$\sigma$ this small, he will no longer find chaotic
solutions. This is confirmed by a glance on eq.~(\ref{stabiloss}): with $\sigma=1$,
the critical parameter $r_c = -5/2$ lies in the
unphysical range $r<0$. If one chooses, instead, some $\sigma > b+1$,
implying $r_c>0$, the window structure in the $r$-dependence is
similar to the case $\sigma = 10$ although the details may be
nontrivially different. But one hardly sees how these differences in
the $r$-behavior for some discrete $\sigma$-values are connected. This
motivated us to vary~$\sigma$ in sufficiently small steps and study the
2-dimensional $(r, \sigma)$-space of control parameters in a larger domain
than has hitherto been done. However, quite apart from such a physical motivation,
the mathematical complexity of this innocent looking set of equations warrants
an investigation of their asymptotic behavior where it has not yet been explored.

In Section~\ref{sec:Two} we present our parameter survey (phase diagrams) and
discuss its main features. Then, in Section~\ref{sec:Attractors} the nature of
two kinds of periodic orbits is described whose stability regions define
the dominant regular features of the large scale phase diagram; an analytic
derivation of these observations is given in Section~\ref{secParameterSpace}.
More complex aspects of the Lorenz dynamics such as bifurcations of periodic orbits, chaos, and
coexistence of attractors, are discussed in Section~\ref{secBifs}. We point out regularities, and
similarities to other well known systems, but do not give detailed explanations of these features.

\section{Two-parameter analysis}
\label{sec:Two}

\begin{figure}
\centering
\includegraphics[width=0.5\textwidth]{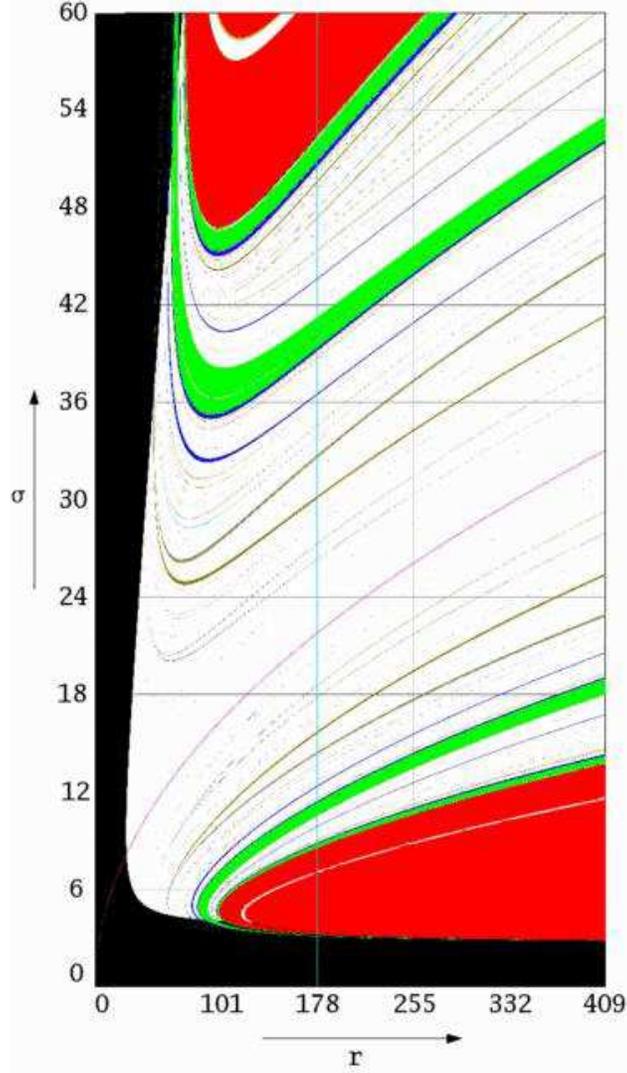}
\caption{\label{fig:r409sigma60}$(r,\sigma)$-phase diagram in the range $0\leq r \leq 409$
         and $0 \leq \sigma \leq 60$.
         Colors indicate the type of attractor reached from the initial condition $(x_0,y_0,z_0) = (0.001,0.001,r-1)$:
         fixed points P$_\pm$ (black),
         orbits of period~2 (red), period~4 (green),
         period~8 (blue), chaotic attractors (white). This picture is a zoom into the lower left corner of
         Fig.~\ref{fig:rsigma2000}. The line $\sigma_{s}:=\sqrt{br}$ in color magenta marks a transition range
         between different global structures for smaller or larger~$\sigma$.
         Further details are discussed in the text, Section~\ref{ssec:Types}.
        }
\end{figure}

\begin{figure}
\centering
\includegraphics[width=\textwidth]{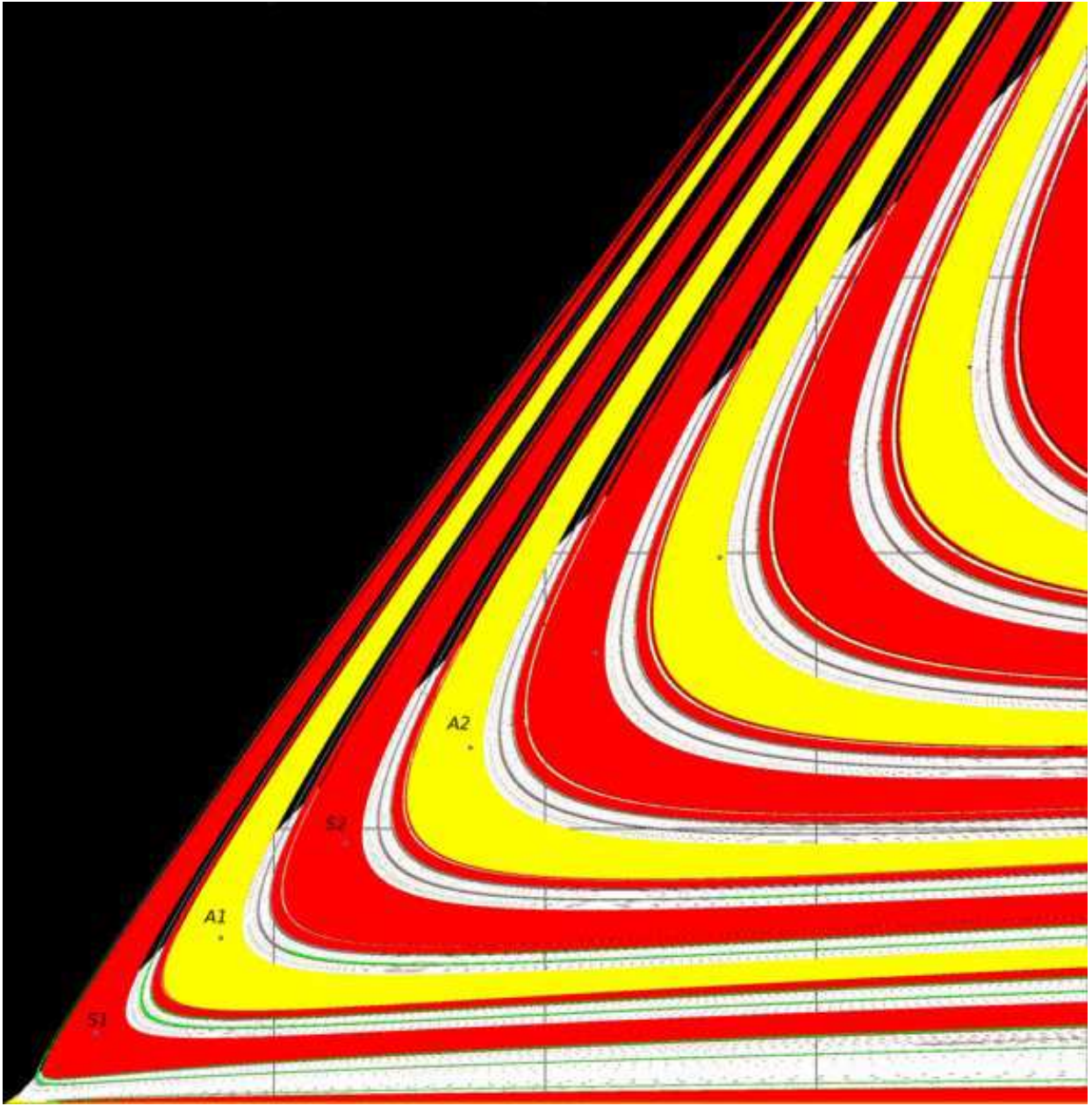}
\caption{\label{fig:rsigma2000} The parameter ranges are $0\leq r,\sigma \leq 2000$.
         The color code is the same as in Fig.~\ref{fig:r409sigma60}. Yellow bands indicate stability
         of orbits with period~1 (not present in the zoomed
         lower left corner, cf.~Fig.~\ref{fig:r409sigma60}). The dots in the red and yellow bands along the line
         $\sigma = 0.75\;r$ are, from bottom left
         to top right, S$_1$, A$_1$, S$_2$, A$_2$, \ldots, defined according to~\eqref{eq:rNfull} for $r_N(m=0.75)$.
         With their upper parts, the bands penetrate well into the range above $\sigma_u = r-28/3$, cf.~\eqref{sigmaul}
         (the line can be identified as the border to the black region through the bands), indicating coexistence of periodic orbits
         with the stable fixed points P$_{\pm}$.
        }
\end{figure}

A glance at Figs.~\ref{fig:r409sigma60} and \ref{fig:rsigma2000} gives an impression of the
bifurcation scheme on large and very large scales of $r$ (abscissa) and $\sigma$. The parameter~$b$,
as mentioned already, is kept at its standard value~$8/3$. Colors indicate the existence of attractors
of a certain type, which does not preclude the coexistence of others. More precisely, the figures are
parameter scans where each point $(r,\sigma)$ is colored according to which type of attractor
is approached from a fixed initial point $(x_0,y_0,z_0)$. Such pictures are called final state diagrams,
or phase diagrams.

The simplest attractors are the fixed points~P$_\pm$, see~(\ref{fixedpoints}).
Their stability is lost along the curve
$r=r_c(\sigma)$, see~(\ref{stabiloss}), or rather by its inverse function $\sigma = \sigma_c(r)$ with
\be
    \label{hyperbola}
    \sigma_c(r) = \frac{r}{2} - \frac{b+3}{2} \pm\frac{1}{2}\sqrt{r^2-2(3b+5)r+(b+3)^2} \ .
\ee
For large~$r$, the upper and lower asymptotes $\sigma_u(r)$ and $\sigma_\ell(r)$ are, respectively,
\be
  \label{sigmaul}
    \sigma_u = r - 2(b+2) \quad \text{and} \quad \sigma_{\ell} = b + 1 \ .
\ee
The next order correction is $\mp (b+1)(2b+4)/r$, describing the approach of the upper
asymptote from below and the lower one from above.

The $(r,\sigma)$-region where the points P$_\pm$ are stable according to this linear analysis
is colored black, the rest is left white, before the other colors are added in Figs.~\ref{fig:r409sigma60} and~\ref{fig:rsigma2000}.
These colors code for the periods of attractors which are determined by the procedure described below:
yellow for orbits of period~1, red for period~2, green for period~4, blue for period~8,
grey for period~3, olive for period~6. %

\begin{figure}[h]


\centering 
\includegraphics[width=0.6\textwidth]{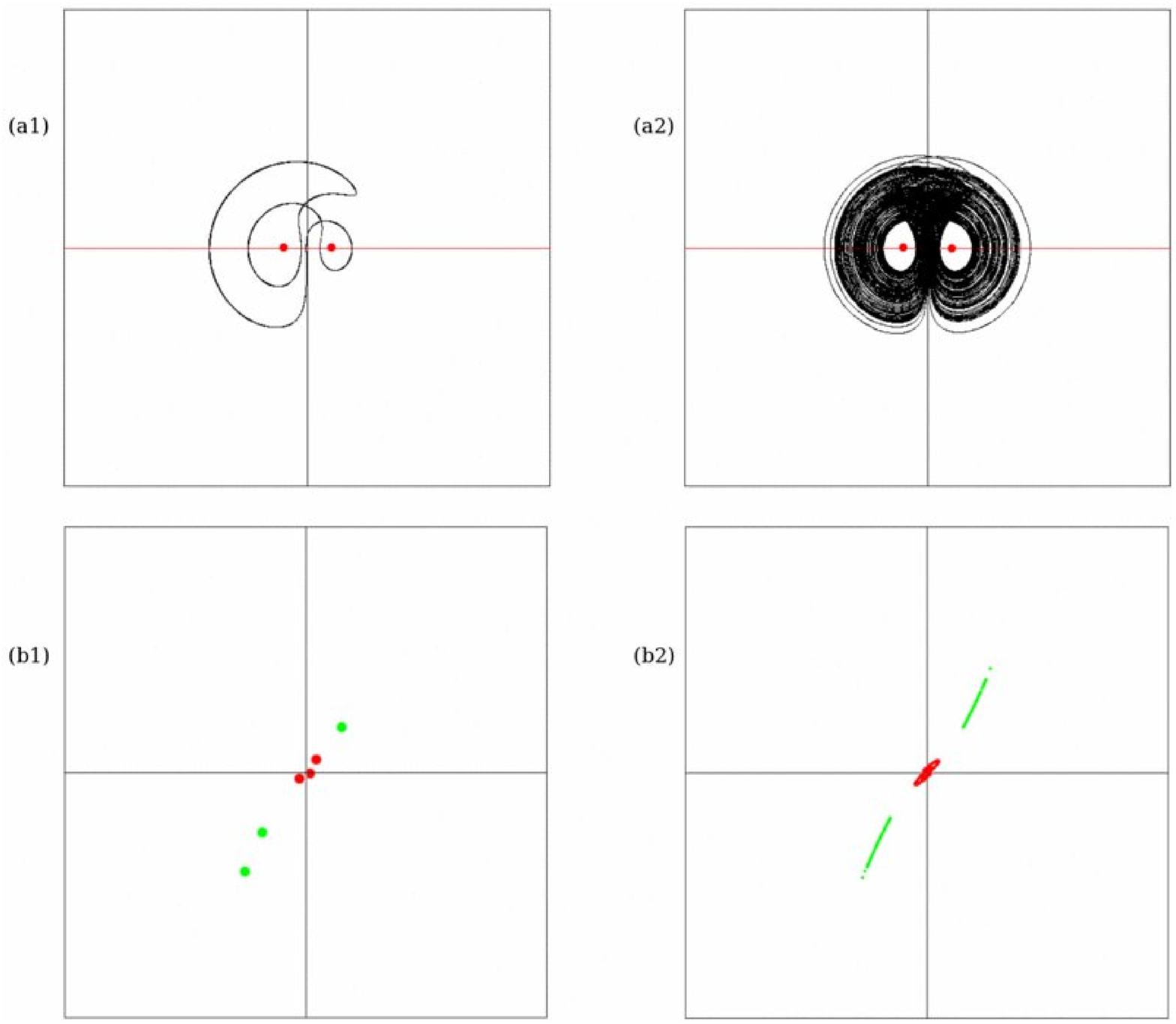}

\caption{\label{Fig:Poinc} (Color online). (a1): $(y,z)$-projection of an asymmetric attractor of period~3 (abscissa~$y$, ordinate~$z$).
         $r=1744$, $\sigma = 110$, initial
         conditions $x_0=y_0=10^{-5}$, $z_0 = r-1$. The fat points are the fixed points~P$_\pm$. (a2): The same for a
         chaotic attractor; $r=1744.9$, $\sigma = 115.2$. (b1) and (b2): Corresponding Poincar\'e sections with the $(x,y)$-plane
         at $z = r-1$; intersection points are grey (green) if from below, black (red) if from above (abscissa~$x$, ordinate~$y$).}
\end{figure}

For each $(r,\sigma)$ on a grid of $610\times 620$
points, an orbit was started with initial conditions $(x_0,y_0,z_0) = (0.001,0.001,r-1)$. A transient regime of 80\,000 iterations
was allowed to elapse before it was assumed that an attractor was reached. The trajectory was then analyzed in terms of
its intersections with the $(x,y)$-plane at $z=r-1$ which is an appropriate  Poincar\'e
surface of section. The period~p of a periodic orbit is defined as half the number of intersection points
of the orbit with that plane
(in one half the trajectory intersects from above, in the other half from below).
Fig.~\ref{Fig:Poinc} gives two examples for illustration.
To determine the period we used a rather simple scheme.
A disk of radius $10^{-3}$ was placed around each point of intersection unless it fell in a disk already generated.
When the number of patches so obtained
remained finite, like in the left part of Fig.~\ref{Fig:Poinc}, the orbit was assumed to be periodic.
When, on the other hand,
the number of disks seemed to grow indefinitely, like in the right part of the figure,
the orbit was considered to be chaotic. In practice we waited for
more than 100 intersections, but only periods up to~8 form clearly identifiable bands on the scales of the figures.
When a periodic orbit was detected, the color of the grid point $(r,\sigma)$
was overwritten with the color of the code given above. Otherwise the original black or white color
was left unchanged. Hence chaotic attractors appear as black or white points.

The method produces two artifacts. First, it cannot identify the fixed point attractors~P$_\pm$; hence,
in the black region it does not distinguish between fixed points and chaotic attractors. Second,
when the relaxation towards an attractor is very slow, the 80\,000 transient iterations may not be sufficient to reach it,
and the method produces more than 100 patches. Hence this situation is not distinguished from chaos. It is observed
in connection with critical slowing down in the neighborhood of a bifurcation of the attractor.
Examples are the white bands in the red regions at bottom and top of Fig.~\ref{fig:r409sigma60},
the narrow seams between red and green regions, and the lines inside the two green bands in the chaotic range.
The nature of the corresponding bifurcations will be elucidated in the following section.

Since only one initial condition is used, the procedure does not determine whether different
initial conditions would lead to different attractors for the chosen parameters $(r,\sigma)$.
The question of coexistence of attractors
will be addressed later, in Sec.~\ref{ssec:Coexistence}.
At this point we can assert already that where a colored band
extends into the black region, like in the lower right and
upper left parts of Fig.~\ref{fig:r409sigma60}, the corresponding periodic orbit coexists with the fixed points~P$_\pm$.

The arrangement of bands in Fig.~\ref{fig:r409sigma60} exhibits a qualitative
structural symmetry with respect to a line which has been added in magenta color. This line,
mathematically defined as $\sigma_s = \sqrt{br}$,  separates
a large-$\sigma$ structure from a kind of mirror image in the low-$\sigma$ part of the
phase diagram. It is conspicuous how every band
below~$\sigma_s(r)$ seems to have a counterpart on the other, upper side.
But, as a glance on Fig.~\ref{fig:rsigma2000} shows, this ``symmetry''
disappears on a larger scale: there is no repetitive structure in the low-$\sigma$ region, in contrast
to the large-$\sigma$ range.

No orbits of period~1 (yellow bands) are observed in the $(r,\sigma)$-range covered in
Fig.~\ref{fig:r409sigma60}. However, in Fig.~\ref{fig:rsigma2000} which covers a much larger $(r, \sigma)$-range,
an onion like repetitive pattern of yellow (period 1) and red (period 2) bands dominates in the upper part.
These bands are separated from each other by chaotic regions which also contain
small bands of higher periods. The bands (above $\sigma_{s}$)
have an hyperbolic shape whose upper branches
seem to have asymptotic slopes of approximately~1.3
while the lower branches will be shown in Section~\ref{secParameterSpace} to have a limiting behavior $\sigma\propto\sqrt{br}$
like the ``bottom line'' $\sigma_s(r)$.
The upper branches penetrate significantly into the region $\sigma > \sigma_u$ where the fixed
points~P$_\pm$ are stable, see eq.~(\ref{sigmaul}). This means there is coexistence
of fixed point attractors and periodic attractors. To generate Fig.~\ref{fig:rsigma2000} the initial
conditions were $x_0 = y_0 = 10^{-5}$ and $z = r+1000$.
Choosing the initial conditions as in Fig.~\ref{fig:r409sigma60},
the corresponding picture (not shown here) looks different in that the greater part of the white bands
of chaos is replaced by colors yellow or red. This reflects the coexistence of chaotic and periodic attractors.

The bands have infinite vertical slopes along the line $\sigma_v(r) \approx 0.75 r$.
On this line, we have marked points S$_1$, A$_1$, S$_2$, A$_2$, $S_3$, ... in the
red (period 2) and yellow (period 1) bands, respectively. The names of these points will also
be taken to denote the corresponding bands
because the attractors of a given band are topologically identical. Their nature will first be
described phenomenologically in the subsequent section, where we shall also see how it changes
from one band to the next.
An analytic explanation of the findings will then be given in Section~\ref{secParameterSpace}.

\section{Attractive periodic orbits}
\label{sec:Attractors}

In this section we explore the nature of the dominant periodic attractors, i.\,e., of the orbits with
periods 1 or 2, by means of numerical and graphical investigation. Their bifurcations into orbits
of higher periods, as well as into strange attractors, will be discussed in Section~\ref{secBifs}.

\begin{figure}
\centering
\includegraphics[width=0.75\textwidth]{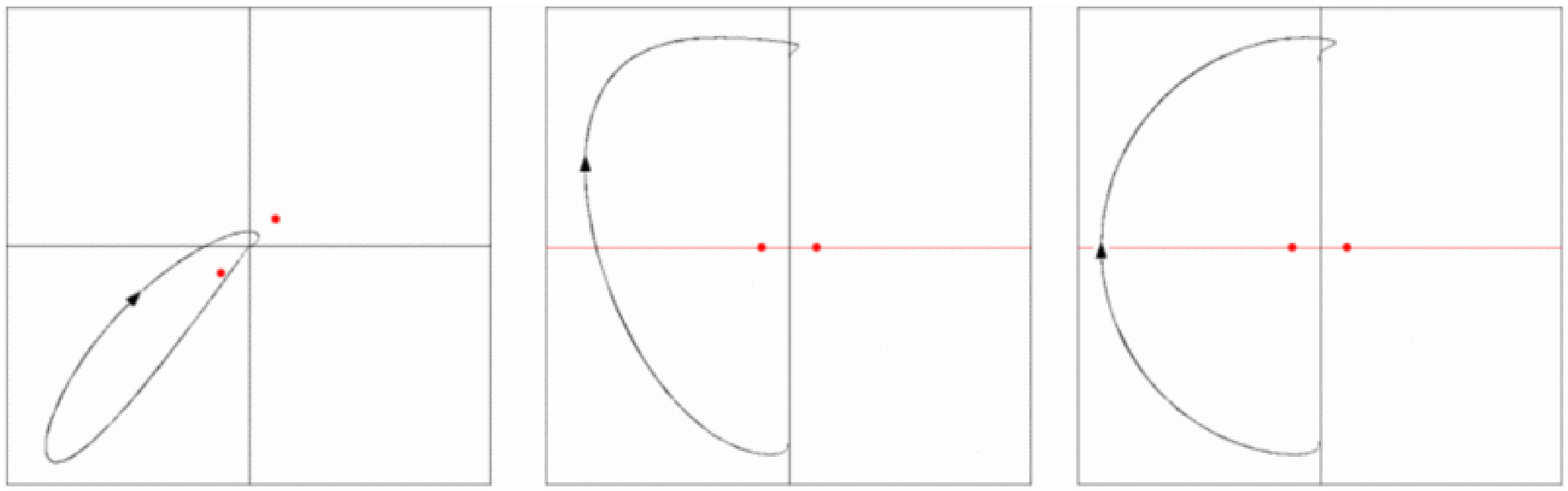}
\caption{\label{fig:AttractorA3}Stable asymmetric periodic orbit corresponding to the parameters
$(r,\sigma)=(1322,992)$ of the point $\text{A}_3$ in Fig.~\ref{fig:rsigma2000}. From left to right,
the attractor is shown in $(x,y)$, $(x,z)$, and $(y,z)$ projection (the abscissa is always mentioned as first coordinate,
the ordinate as second). The dots mark the projections of
the fixed points $\text{P}_{\pm}$, see~(\ref{fixedpoints}).
The arrows indicate the direction of flow; the decrease of~$z$ takes place very close to the $z$-axis.
The attractor coexists with its $\Sigma$-partner
which is obtained by reflection of $x$ and~$y$. Together these two attractors
look like the symmetric attractor in Fig.~\ref{fig:AttractorS4}.
The scales are $-530<x,y,z-r<530$.}
\end{figure}

\begin{figure}
\centering
\includegraphics[width=0.75\textwidth]{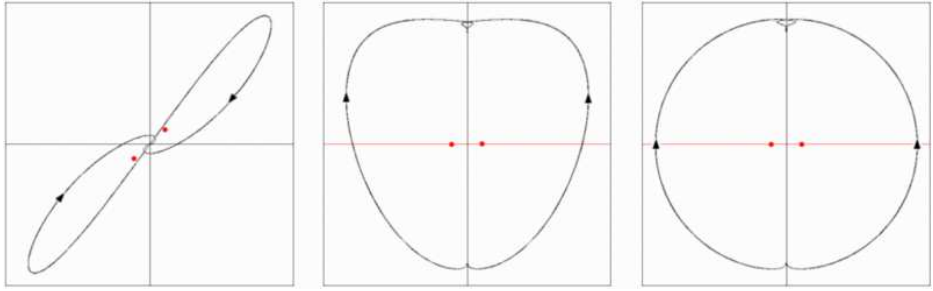}
\caption{\label{fig:AttractorS4}Stable symmetric periodic orbit corresponding to $(r,\sigma)=(1552,1164)$, the point
$\text{S}_4$ in Fig.~\ref{fig:rsigma2000}. Same projections as in Fig.~\ref{fig:AttractorA3}.
The scales are $-620<x,y,z-r<620$.}
\end{figure}

\begin{figure}
\centering
\includegraphics[width=0.75\textwidth]{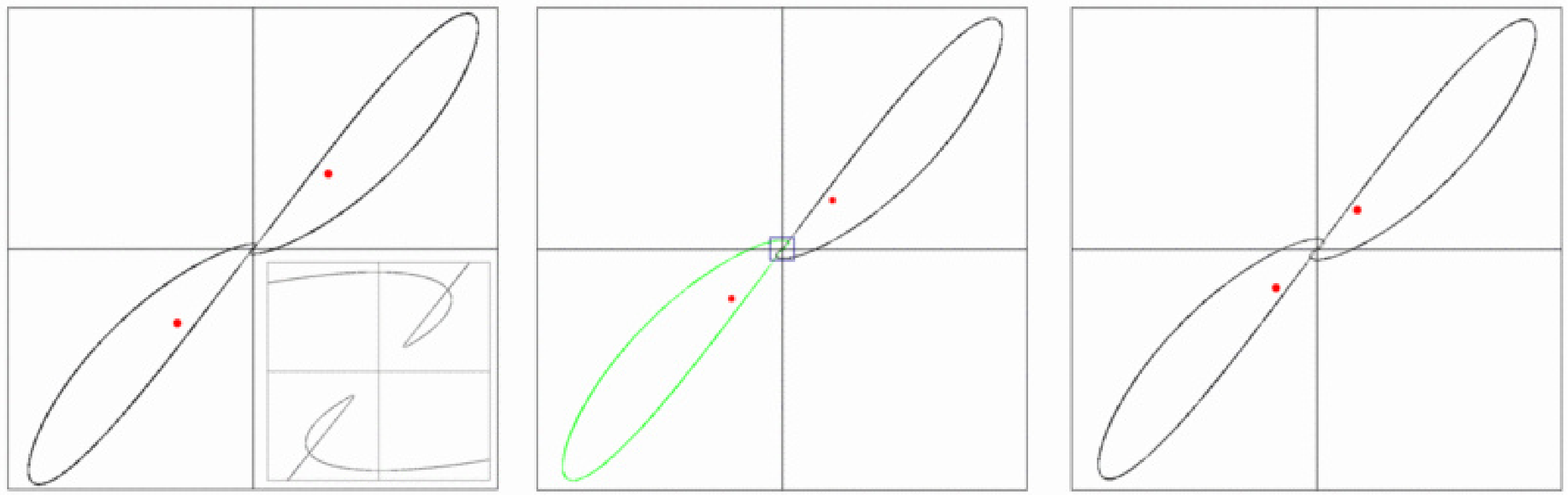}
\caption{\label{fig:Winding} (Color online). 
$(x,y)$-projections of (from left to right) the attractors of S$_1$, A$_1$, S$_2$.
In the case of A$_1$ the two asymmetric orbits are shown in different color. 
The insert in the left picture is a blow-up
of the neighborhood of the origin.
The $(r,\sigma)$ values are S$_1$ = (172,130), A$_1$ = (402,302), S$_2$ = (632,474); the scales are
$-0.4\,r<x,y< 0.4\,r$.
}
\end{figure}

To begin with, consider the attractors corresponding to the $(r,\sigma)$-values of A$_3$ in
Fig.~\ref{fig:AttractorA3} and of S$_4$ in Fig.~\ref{fig:AttractorS4},
in the three standard projections. Comparison with the S$_1$, A$_1$, S$_2$, and A$_2$
attractors (see Figs.~\ref{fig:Winding} and~\ref{fig:rxbifdia}) shows that on the scales used, $-0.4r < x,y,z-r < 0.4r$,
all A$_n$ attractors look alike, and so do the orbits of~S$_n$. Furthermore, the attractor of the parameter
point S$_n$ (for given $n$) looks like a combination of the asymmetric orbits of A$_n$ and its $\Sigma$-mirror
image~$\Sigma\text{A}_n$.

\begin{figure}
\centering
\includegraphics[width=0.8\textwidth]{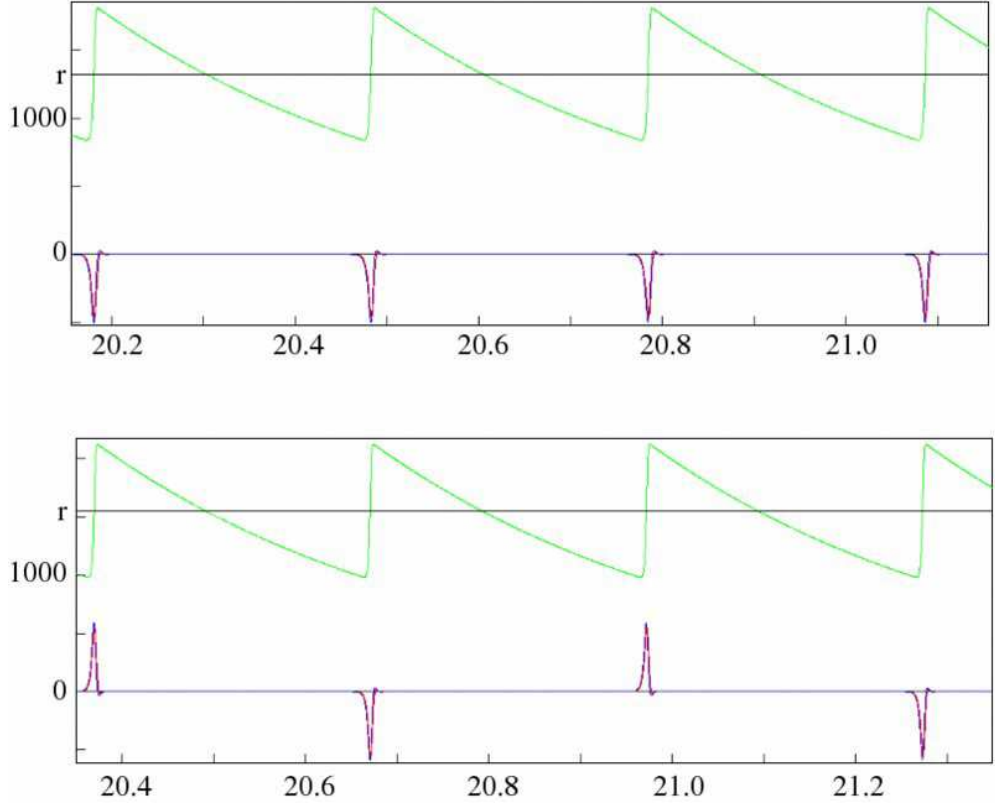}
\caption{\label{fig:A3S4intime} (Color online).  
Time development of $x$, $y$, and $z$ within one unit
of time~$t$ (abscissa), for the attractors of $\text{A}_3$ (top) and of $\text{S}_4$ (bottom).
The upper parts show the $z$-coordinates, the lower parts $x$ and~$y$ which are nearly indistinguishable on this time scale.
The vertical scale is $-0.4\,r<x,y,z-r< 0.4\,r$.}
\end{figure}

The time course of the variables $x$, $y$, and $z$ is plotted in Fig.~\ref{fig:A3S4intime}, both for the
A$_3$ attractor (top) and the S$_4$ one (bottom).

The combined picture shows the typical behavior of a relaxation oscillator.
Two different regimes can clearly be distinguished: First, the variable $z$ decreases slowly from an upper
value $r(1+\hat\rho)$ to a lower value $r(1-\hat\rho)$, with $\hat\rho \approx 0.35$, while $x$ and $y$ are virtually zero.
Then, in a very fast process, $x$ and $y$ start out on a large excursion with amplitude $r\hat\rho$, while $z$ jumps back to
its upper value. The $(y,z)$-projection exhibits an almost perfect half circle; the $(x,y)$-projection shows that
during the excursion, $x$ lags somewhat behind~$y$.

\begin{figure}
\centering
\includegraphics[width=0.75\textwidth]{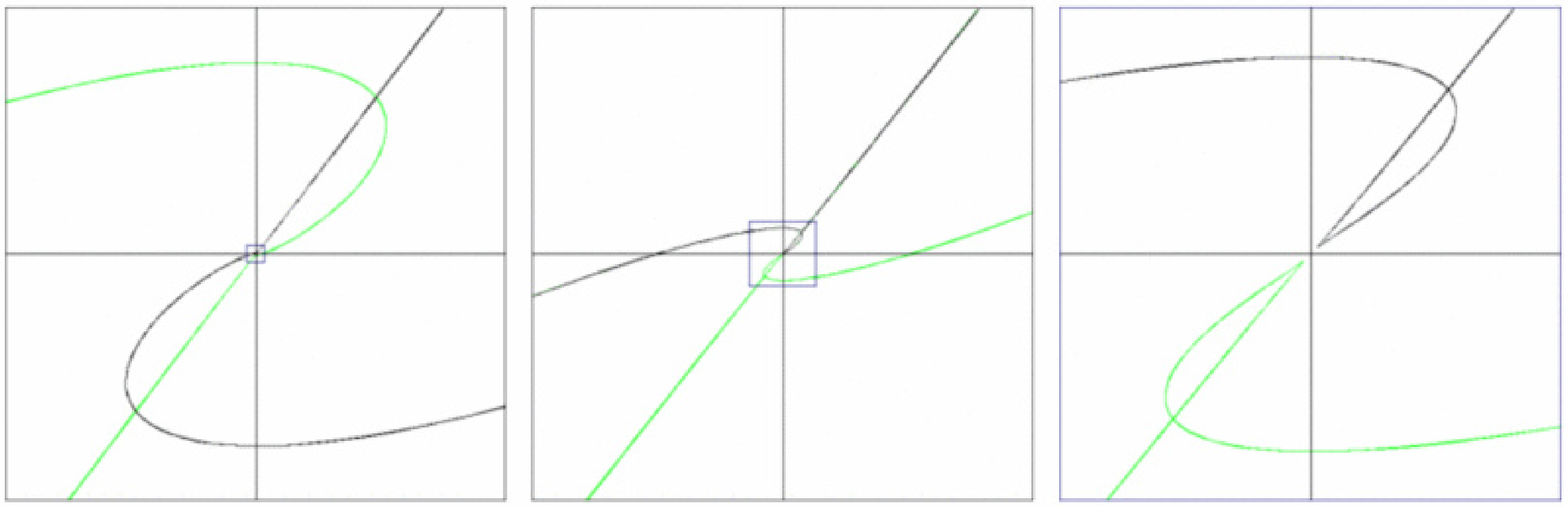}
\caption{\label{fig:A1Blowup}  (Color online). 
Successive blow-ups of the box in the middle picture of Fig.~\ref{fig:Winding}.
The sequence demonstrates that the two orbits of A$_1$ do not reach the origin; rather each of them winds once around the $z$-axis.
}
\end{figure}

\begin{figure}
\centering
\includegraphics[width=0.8\textwidth]{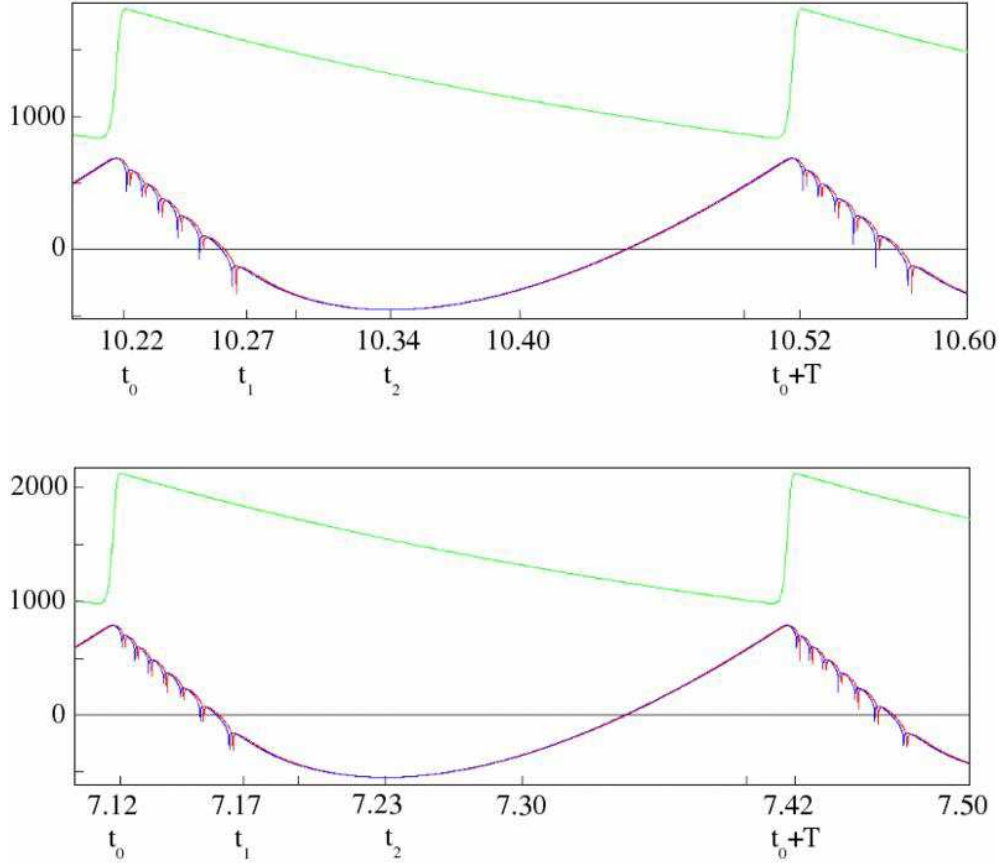}
\caption{\label{fig:A3S4Logintime}  (Color online). 
Time development of $\log |x|$ (red),  $\log |y|$ (blue), and above them $z$ in grey (green) 
within 0.4 time units,
for the attractors of $\text{A}_3$ (top) and of $\text{S}_4$ (bottom). The vertical scales are the same as in
Fig.~\ref{fig:A3S4intime}. $x$~and~$y$ are plotted as $30 \log |x,y| + 600$.
Thus the $x$- and $y$-zeros appear as logarithmic singularities.}
\end{figure}

From Figs.~\ref{fig:AttractorA3}-\ref{fig:A3S4intime} it is not obvious what happens during the slow phase
where the orbits move down along the $z$-axis. This question is clarified by the sequence of $(x,y)$-projections
in Figs.~\ref{fig:Winding} and~\ref{fig:A1Blowup}.
The individual pictures in Fig.~\ref{fig:Winding} show the attractors of S$_1$, A$_1$, S$_2$; the other projections
would be similar to those in Figs.~\ref{fig:AttractorA3} and~\ref{fig:AttractorS4}. Taken together, the pictures make it clear
that the orbits wind around the $z$-axis on spirals with decreasing $R^2 = x^2+y^2$. Coming from large values
of $R$, near the maximum value of~$z$, they enter a close neighborhood of the $z$-axis when they pass a point
where $y=0$. After a quarter turn in clockwise direction (looking down the $z$-axis) there comes a point where~$x=0$,
at smaller values of~$z$ and~$R$.
What happens next depends on the index $n$ of S$_n$ or A$_n$. In the case of S$_1$ the orbit quickly leaves the
$(x,y)$-origin, cf.~the inset in the left part of Fig.~\ref{fig:Winding}; it performs
a large loop and returns to the $(x,y)$-origin region from the other side, but again near the maximum of~$z$.
All in all it encircles the $z$-axis once, its winding number~$W$ with respect to the $z$-axis being equal to
the number $N=1$ of zeros in $y$ (and correspondingly in $x$) upon each close encounter. In the case
of A$_1$, the orbit stays longer near the $z$-axis, with a second zero of~$y$ and also of~$x$, each with
smaller values of~$R$ (and~$z$); this is illustrated by the sequence of zooms in Fig.~\ref{fig:A1Blowup}.
Its winding number with respect to the $z$-axis is also~$W=1$, but now there
are two of these orbits, shown in different colors. To keep track of the winding number~$W$ for increasing band
label~$n$, we may focus on the number~$N$ of zeros of~$y$ (or $x)$ upon close encounter with the
$z$-axis. Detailed observation shows that this number is
\be
 \label{Windingnumbers}
  W(\text{S}_n) = N(\text{S}_n) = 2n-1  \qquad \text{and} \qquad W(\text{A}_n)=\halb N(\text{A}_n) = n  .
\ee
Fig.~\ref{fig:A3S4Logintime} shows a logarithmic plot of $|x|$ and $|y|$ for A$_3$ ($N=6, W=3$) and
S$_4$ ($N=7, W=7$).
We observe three different regimes of time dependence. Soon after $z$ reaches its peak value, at a time $t_0$,
the first zeros of $y$ and $x$ occur, visible as logarithmic singularities.
The first regime of oscillatory exponential decay of $x$ and~$y$ ends at a time $t_1$. In the second phase
there are no further $x$- or $y$-zeros, but the $z$-decay continues with deceleration up to a time $t_2$
where the distance~$R$ of the orbit from the $z$-axis reaches a minimum.
The third regime is one of explosive, i.\,e., faster than exponential increase to the original values
of $x, y$ and~$z$.

The spiraling behavior in the first regime seems reminiscent of ``Shilnikov Chaos''~\cite{Shilnikov98}.
However, in this case all the relevant dynamics stays far away from the fixed point.
The invariant line $x=y=0$ is part of the stable manifold of the origin,
but this is not important for our arguments. The main ingredient is that it
is attractive for $z > r$ (i.\,e.~up to time $t_2$) and repulsive for $0 < z < r$.
Long before the orbit gets close to the origin it leaves the $z$-axis along
its unstable eigenvector, with $x - y \approx 0$.

The global structure of the $(r,\sigma)$ parameter space, as exhibited in Fig.~\ref{fig:rsigma2000}, is
obviously related to the existence and stability of such orbits, with winding numbers $W(\text{A}_n), W(\text{S}_n)$
increasing as $r$ and~$\sigma$ (and thereby~$n$) grow. The observations to be
explained in the subsequent theoretical section are the following:
\begin{enumerate}
 \item Repetition of bands:
Along the line $\sigma=mr$ with a fixed ratio $m$, there exists a period~$q$ in~$r$ (and consequently
$mq$ in~$\sigma$) such that if an attractor of type A$_n$ (or S$_n$) exists for values $(r_n,\sigma_n)$,
then a corresponding attractor of type A$_{n+N}$ (or S$_{n+N}$) exists for $(r_n + Nq, \sigma_n + Nmq)$.
 \item $m$-dependence of $r$-periods~$q$:
From the shape of the bands it may be inferred that $q(m)$ has a minimum near $m=0.75$.
This means, the periodic repetition of the bands is fastest (in~$r$) along the line $\sigma=0.75 r$.
 \item Size scaling of dominant orbits with~$r$:
 The size of these orbits, i.\,e., the radius $\hat\rho$ of their $(y,z)$-projections measured in units of~$r$, is
 independent of~$r$: $\hat\rho \approx \const$ for fixed~$m$.
\end{enumerate}
Thus far we have not addressed the question of how $\hat\rho$ depends on~$m$, so let us see next what theory has to tell;
the result is presented in Fig.~\ref{fig:rhomegavsm} where $\rho = \hat\rho/m$ is plotted as a function of~$m$.
Of course, the above statements only hold for large~$r$ and~$\sigma$, in practice for $r\gtrsim 100$
and $\sigma$ in the bands above~$\sigma_s=\sqrt{br}$.

\section{Analysis of the parameter space structure}
\label{secParameterSpace}

We begin with a discussion of appropriate scalings of the entire set of Lorenz equations~(\ref{lorenz}) when
both parameters~$r$ and~$\sigma$ become large. Their asymptotic behavior with $r\rightarrow \infty$ depends
on how $\sigma$ is assumed to change simultaneously. Robbins in~\cite{Ro79} and also Sparrow~\cite{Sp82}
considered the case $\sigma=\const$. Here we study the behavior of the Lorenz equations with Prandtl numbers
increasing as $\sigma \propto r^s$, with some fixed exponent $s>0$ to guarantee increasing~$\sigma$, but $s \leq 1$
for $\sigma$ to stay below the upper stability border $\sigma_u(r) = r - 28/3$.

If r and $\sigma$ become very large,
the motion in~$x$ and~$y$ is expected to become very fast because $\d /\d t$ in the equations of motion becomes large.
Thus we introduce a rescaled, stretched time~$\tau$
by defining $\d\tau = \varepsilon^{-1} \d t$ with a scaling factor $\varepsilon^{-1} \propto r^{e}$, $e>0$.

Let us now also rescale the variables $x$ and $y$ which measure, respectively, the amplitudes of the role velocity and temperature.
As for the rescaling of~$z$ a glance on the second Lorenz equation shows that $y$ is driven by $(r-z)$. This fact is reflected
in the amplitudes~$z$ of the numerical trajectories, or in the behavior of the fixed points~P$_\pm$,
both as functions of~$r$. Thus we define
\be
\label{rescaling}
x = \alpha \xi, \qquad y = \beta\eta, \qquad z-r = \gamma\zeta .
\ee
The Lorenz equations then read, with $\xi'=\d\xi/\d\tau$ etc.,
\be
\label{lorenzrescaled}
\xi'=-\varepsilon\sigma\xi+\varepsilon\sigma\frac{\beta}{\alpha}\eta, \qquad
\eta'=-\varepsilon\eta-\varepsilon\gamma\frac{\alpha}{\beta}\zeta\xi, \qquad
\zeta'=-\varepsilon b (\zeta+\frac{r}{\gamma})+\varepsilon \frac{\alpha\beta}{\gamma}\xi\eta  .
\ee
The freedom of choosing scaling parameters may now be used to normalize to unity the coefficients of
the driving term in the first equation, and of the two coupling terms $\propto \zeta\xi$ and $\propto \xi\eta$, which leads to
\be
\label{scalerelations}
\varepsilon\sigma\beta=\alpha , \qquad \varepsilon\gamma\alpha=\beta , \qquad \varepsilon\alpha\beta=\gamma .
\ee
These equations allow us to express $\alpha$, $\beta$, and $\gamma$ in terms of $\sigma$ and $\varepsilon$.
We find first $\varepsilon^2\alpha^2=1$, thus $\alpha=\pm\varepsilon^{-1}$, then
$\beta=\pm\varepsilon^{-2}\sigma^{-1}$, and finally $\gamma=\varepsilon^{-2}\sigma^{-1}$. The $\pm$ signs
may be interpreted as corresponding to the symmetry~$\Sigma$ which is clearly preserved under the rescaling.
We can therefore restrict ourselves to the plus sign and have
\be
\label{factors}
\alpha=\varepsilon^{-1}\propto r^e , \qquad \beta= \gamma = \varepsilon^{-2}\sigma^{-1}\propto r^{2e-s} .
\ee
The asymptotic behavior will depend on how the exponent~$e$ compares to the other exponent~$2e-s$. There appear to be three
distinct possibilities. One is that $\beta$ and $\gamma$ grow faster than~$\alpha$, i.\,e., $2e-s > e$ or $e>s$.
The second is that all three variables $x, y, z$ scale with the same factor, $2e -s = e$ or $e=s$. The third possibility
$e < s$ may be excluded with the argument that then the relaxation term $-\varepsilon\sigma\xi$ in the $\xi$-equation would
diverge.

A representative case of the first possibility is $s=0$ and $e = \halb$, meaning $\sigma = \const$ and
$\varepsilon \propto 1/\sqrt{r} \rightarrow 0$. This was the choice of
Robbins~\cite{Ro79} and Sparrow~\cite{Sp82}. All dissipative terms vanish in this limit, and the equations reduce to
\be
 \label{Sparrow}
  \xi'   = \eta \, , \qquad
  \eta'  = -\zeta\xi \, , \qquad
  \zeta' = \xi\eta  \, ,  \qquad (\sigma = \text{const}, \ \varepsilon = 1/\sqrt{r}) .
\ee
This system has the two integrals $A = \frac12\xi^2-\zeta$ and $B^2=\eta^2+\zeta^2$; it
is equivalent to the equations of a pendulum with length~$B$ and energy~$A$. The values of the integrals $A$ and~$B$
have to be determined from the $\varepsilon$-correction to~(\ref{Sparrow}), i.\,e., by considering their slow dynamics.
In~\cite{Ro79,Sp82} it was shown how this can be done in terms of elliptic integrals. In that sense, this case has been solved.

Let us therefore concentrate on the second possibility $e=s$. Then $\varepsilon\sigma$ is independent of~$r$, and we can take
$\varepsilon\sigma = 1$. All scaling factors can now be expressed in terms of the Prandtl number~$\sigma$.
The term $-\varepsilon\eta$ in the second Lorenz equation may be neglected as vanishingly small in comparison to with~$\zeta\xi$.
Hence the rescaled eqs.~(\ref{lorenzrescaled}) become
\be
\label{lorenzrescaled2}
\xi'=-\xi+\eta , \qquad \eta'=-\zeta\xi , \qquad
\zeta' = \xi\eta - \frac{b}{\sigma}\zeta - \frac{br}{\sigma^2} , \qquad
     \text{($e=s$ and $\sigma = r^s$)}\ .
\ee
But what about the two last terms in the $\zeta$-equation? Their order of magnitude
is $b r^{-s}$ and $b r^{1-2s}$, respectively.
Again, there are three possibilities: \\
Case A. With the choice $1-2s=0$, or $s=1/2$, the second term will tend to a constant and dominate the decreasing first term.
We can write $\sigma = w\sqrt{br}$ with some constant~$w$. The set of equations becomes
\be
\label{case1/2}
\xi'=-\xi+\eta, \qquad \eta'=-\zeta\xi, \qquad \zeta'=\xi\eta-w^{-2} \qquad
     \text{($e=s=\halb$ and $\sigma = w\sqrt{br}$)}\ .
\ee
When $s$ varies between $1/2$ and $1$, the same equations apply because the last term in the
$\zeta$-equation of~(\ref{lorenzrescaled2}) dominates the second. Note that no comparison can be made between these two
terms and $\xi\eta$. \\
Case B. When $s=1$, the two last terms in the $\zeta$-equation are of the same order of magnitude~$r^{-1}$.
We write $\sigma = mr$ and obtain the set of equations
\be
\label{case1}
\xi'=-\xi+\eta, \qquad \eta'=-\zeta\xi, \qquad \zeta'=\xi\eta-\frac{b}{mr}\Bigl(\zeta+\frac{1}{m}\Bigr)  \qquad
     \text{($e=s=1$ and $\sigma = mr$)}\ .
\ee
Case C. When $s>1$, $\sigma$ grows faster than~$r$, so we leave the range where $\sigma<\sigma_u(r)$.
Formally it is found that both terms besides $\xi\eta$ in the $\zeta$-equation may be neglected, but
in contrast to eqs.~(\ref{Sparrow}) there remains a dissipative term in the $\xi$-equation.
The constant of motion $B^2 = \eta^2 + \zeta^2$ constrains the solution to a cylinder.
On this cylinder the equation reduces to the damped pendulum, and therefore almost all
initial conditions decay towards the stable equilibrium on the respective cylinder.

We thus have to discuss the asymptotic regimes A~and~B, and do this separately. To our knowledge, the two sets of eqs.~(\ref
{case1/2}) and (\ref{case1}) have not been analyzed so far. As we shall see, their solutions lead to an
understanding of the phase space structure displayed in Fig.~\ref{fig:rsigma2000} and of the numerical
findings in Section~\ref{sec:Attractors}.

\subsection{The limit of large $r$ with $\sigma = w\sqrt{br}$}
\label{ssecw}

We start with the simpler case of the asymptotic Lorenz equations, the set of the rescaled eqs.~(\ref{case1/2}).
Of all the complex dynamics that it contains, we only want to describe
periodic orbits of the simplest kind. From our numerical studies we anticipate that these orbits have two
clearly distinct phases. There is slow motion down along the $z$-axis, and a fast return to the
original large value of~$z$ via a large loop away from the $z$-axis.

The fast motion is readily described when we neglect the constant term in the $\zeta$-equation of
(\ref{case1/2}). Then, combining the equations for $\eta$ and $\zeta$,
we obtain $\eta\d\eta = - \zeta\d\zeta$ which after integration gives the circles
\be \label{eqn:etazetaloop}
    \eta^2 + \zeta^2 = \rho^2
\ee
with as yet undetermined radius~$\rho$.
A convenient parametrization of these circles is found with a rescaling of time,
defining $s$ as an arc length via $\d s = \xi(\tau)\d \tau$. The
equations for $\eta(s)$ and $\zeta(s)$ are thereby linearized, with solutions
\be \label{eqn:etazetaofs}
    \eta(s) = \rho\sin s, \qquad \zeta(s) = - \rho\cos s \, .
\ee
This explains the observation that $\zeta$ oscillates between
two equal values of different sign, namely $\pm \rho$.

The equation for $\xi$,
\be \label{eqn:Loopxi}
    \frac{\d\xi}{\d \tau} = -(\xi - \eta) \qquad \text{or} \qquad
    \frac{\d\xi}{\d s} = - \Bigl( 1-\frac{\eta(s)}{\xi(s)} \Bigr)
\ee
does not appear to be analytically solvable, but as long as~$\xi$ is not too small, it describes a
relaxation of $\xi(s)$ towards $\eta(s)$. Hence, with a delay in~$\tau$ of the order of $1$, $\xi$
follows~$\eta$ on its excursion
to values of the order of~$\rho$.

Consider now the slow motion. It is given by the exact dynamics on the stable manifold $\xi = \eta = 0 $
of the fixed point $(0, 0, 0)$. The solution with initial point $\zeta(0) = \rho$ is
$\zeta(\tau) = \rho - \tau/w^2$
and reaches the lower value $-\rho$ at time $T = 2 \rho w^2$. Linearizing the set of
equations~(\ref{case1/2}) around this motion, we obtain
\be
       \xi'   =  - \xi  + \eta  \, , \qquad
      \eta'  = -(\rho - \tau/w^2)\xi \, .
\ee
The $\tau$-dependent eigenvalues are $-\frac12 \pm \sqrt{\Delta(\tau)}$ with discriminant $\Delta = \frac14-\rho + \tau/w^2$.
Three $\tau$-regimes may be identified. Assuming $\rho > 1/4$,
the discriminant is initially negative; the motion is oscillatory and decaying according to $\text{e}^{-\tau/2}$.
This regime ends at the time $\tau_1$ defined by $\Delta(\tau_1) = 0$, or $\tau_1 = w^2(\rho-\frac14)$.
In the second regime, the motion continues to be decaying but is no longer oscillatory; this is the case
between $\tau_1$ and $\tau_2 = w^2\rho = T/2$ where $\zeta(\tau_2)=0$.
The third regime, between $\tau_2$ and $T$, sees $\xi$ and $\eta$ increasing at a growing rate.

To proceed with the $(\xi,\eta)$-equations, we introduce a new variable $X$ by separating the
exponentially decaying part from $\xi$.
With $\xi(\tau) = e^{-\tau/2} X(\tau)$, the two linear equations of first order are transformed into
\be \label{eqn:XAi}
   X''  - \Delta(\tau) X  = 0 \,, \qquad \text{or} \qquad \tilde X'' - \frac{\tilde\tau}{w^2} \tilde X = 0
\ee
where $\tilde\tau = \tau-\tau_1$, and $\tilde X(\tilde\tau) = X(\tau -\tau_1)$ lives on the
interval $-\tau_1 < \tilde\tau < T - \tau_1$.
This equation is the well known Airy equation. Its general solution is a superposition of the two Airy functions
$\text{Ai}(z)$ and $\text{Bi}(z)$ where $z=w^{-2/3}\tilde\tau$.  Both solutions oscillate
for $z<0$. For $z>0$ the solution Ai decays whereas Bi grows exponentially. We are
interested in the solution $\tilde X(\tilde\tau) = \text{Bi}(z)$, or
\be \label{eqn:xiAiSol}
  \xi(\tilde\tau) = \text{e}^{-\tau_1/2}\text{e}^{-\tilde\tau/2}\text{Bi}(w^{-2/3}\tilde\tau) \,.
\ee
with $-\tau_1 < \tilde \tau < T-\tau_1$.
It may be checked that $\eta(\tilde\tau) = e^{-\tau/2} Y(\tau)$ is governed by the same Airy equation and thus,
up to a phase shift, has the same solution. An example of the function Bi is shown in the left
part of Fig.~\ref{fig:Airyplots}, where the parameters are those of~S$_4$ (the plots are made
on the basis of eqs.~(\ref{eqn:xiSol})-(\ref{eq:rhovsm})).
The corresponding~$\xi(\tilde\tau)$ is shown in the right part of the picture, in logarithmic representation.

From here it is straightforward to determine the value of~$\rho$, and the periodicity of the pattern of bands.

We know from the analysis of the outer part of the attractor, see~(\ref{eqn:etazetaofs}), that the solution starts,
for $\tau=0$ or $\tilde\tau = -\tau_1$, at about the same values~$\xi$ and $\eta$ at which it ends
for $\tau= T$:
\be
 \label{rhoEquation}
           \xi(0) = \xi(T) \, .
\ee
Using the asymptotic behavior of Bi for large negative
and positive arguments,
\be
    \label{AiryAsymp}
\begin{split}
  \text{Bi}(-|z|) &\approx \frac{1}{\sqrt{\pi}|z|^{1/4}} \cos \bigl( \textstyle\frac{2}{3}|z|^{3/2} + \frac{\pi}{4}\bigr)\ , \\
  \text{Bi}(|z|)  &\approx \frac{1}{\sqrt{\pi}|z|^{1/4}} \text{exp} \{ \textstyle\frac{2}{3}|z|^{3/2} \} \ ,
\end{split}
\ee
and the fact that the exponential terms dominate the algebraic factors by orders of magnitude,
we simply equate the exponents on the two sides of (\ref{rhoEquation}). For $\tilde\tau = -\tau_1$ eq.~(\ref{eqn:xiAiSol})
gives $\xi(0) = O(1)$ (zero exponent); for $\tilde\tau = T-\tau_1$, we add up the exponents in eqs.~(\ref{eqn:xiAiSol})
and~(\ref{AiryAsymp}), with $z=w^{-2/3}(T-\tau_1)$, and
obtain
\begin{equation}
 \label{expo}
  0 = -\frac{\tau_1}{2} - \frac{T-\tau_1}{2} + \frac23 \bigl|w^{-2/3}(T-\tau_1)\bigr|^{3/2} =
   -\frac12 T + \frac23 \frac{1}{w} (T - \tau_1)^{3/2}  \, .
\end{equation}
Inserting $T = 2 \rho w^2$ and $\tau_1 = w^2\,(\rho - \frac{1}{4})$ we obtain
\be
    \label{rhovalue}
   144 \rho^2 = (1 + 4\rho)^3 \, .
\ee
The relevant solution is $\rho = 1.35... =: \rho_0$.

To determine the values $w = w_N$ where the orbits have a certain number~$N$ of $\xi$- and $\eta$-zeros
at each close encounter with the $z$-axis, we use the asymptotic behavior of the Airy function $\text{Bi}(-|z|)$
as given in~(\ref{AiryAsymp}). Consider first the symmetric orbits~S$_n$ as they develop from time
$\tilde\tau=-\tau_1$ to $T-\tau_1$. The coordinates $\xi$ and $\eta$ come in from large values (positive
or negative) and go out to large values of the opposite sign. This means the phase shift between
$\xi(\tilde\tau=-\tau_1)$ and $\xi(\tilde\tau = T-\tau_1)$ is~$\pi$; the cosine performs an odd number of
half oscillations. For the asymmetric orbits~A$_n$, on the other hand, the phase shift is a multiple
of~$2\pi$. Taken together, this leads to the ``quantum condition''
\be
 \label{wN1}
   \textstyle \frac23\frac{1}{w}\tau_1^{3/2} + \frac{\pi}{4} = N \pi \, ,
\ee
where $N=1$ corresponds to the first symmetric and $N=2$ to the first asymmetric band.
With $\rho = \rho_0$ we get
\be
    \label{wN}
    w_N = 1.02 \sqrt{4N-1}\, , \qquad N =1,2,...\, .
\ee

To wrap it up, the results for this regime of large~$r$ are the following.
The $N$-th band is centered around the line $\sigma_N = w_N\sqrt{br}$;
the size of the corresponding attracting periodic orbits in the original variables $(x,y,z)$ is $\rho_0\sigma_N$.
But note this is only asymptotically true and can hardly be seen in the diagram of Fig.~\ref{fig:rsigma2000}
except in the lower right corner where the S$_1$-band is just bending over to follow the $\sqrt{r}$ behavior.
$r=2000$ gives the prediction $\sigma_1 = 74.5 \sqrt{3} = 129$ which
lies at the lower border of the S$_1$-band.
The bands corresponding to higher values of~$N$ have not yet
entered this asymptotic regime.

\subsection{The limit of large $r$ with $\sigma = mr$}
\label{ssecm}

Let us now study the set of equations (\ref{case1})
\[
  \xi'=-\xi+\eta, \qquad \eta'=-\zeta\xi, \qquad \zeta'=\xi\eta-\frac{b}{mr}\Bigl(\zeta+\frac{1}{m}\Bigr)   ,
\]
where $m = \sigma/r$ is the constant parameter.
Much of the analysis is similar as in the previous subsection, but the explicit occurrence of~$r$ in the
equations adds complications. The general strategy is the same. For the outer, large $\xi$ and $\eta$ parts of
the solution we neglect the term
$\propto r^{-1}$ in the $\zeta$-equation and get identical results as in (\ref{eqn:etazetaloop})-(\ref{eqn:Loopxi}).

The slow motion along the $z$-axis is more complicated even though the $\zeta$-equation, neglecting the
$\xi \eta$-term, decouples again from the rest:
its solution is no longer linear in $\tau$. With the initial condition $\zeta(0) = \rho$ the approximate
$\zeta'$-equation is solved by
\be
\label{zetaoftau}
    \zeta(\tau) =  (\rho + {\textstyle\frac1m})\text{e}^{-(b/mr)\tau} - {\textstyle\frac1m},
\ee
and the time $T$ needed to evolve from $\rho$ to $-\rho$ is 
\be \label{eqn:T}
   T =  \frac{mr}{b} \log \frac{1+m\rho}{1-m\rho} \, .
\ee
Using this slow exponential decrease of $\zeta$ (with the rate $\propto r^{-1}$) in the
equation $\eta'=-\zeta(\tau)\xi$,
we have again a time dependent linear system of equations for $\xi$ and $\eta$.
Its eigenvalues at fixed time are
\be
 \label{xietaEigenvalue}
  \lambda = -\textstyle\frac{1}{2} \pm \sqrt{\Delta} \qquad \text{with} \qquad
   \Delta = \textstyle \frac{1}{4} -\zeta(\tau) = \frac{1}{4}+ \frac1m - (\rho +
\frac1m ) \displaystyle \text{e}^{-(b/mr)\tau} \, .
\ee
The discriminant $\Delta$ is negative in the $\tau$-range  $0<\tau<\tau_1$ where $\tau_1$ is defined by
the condition $\Delta(\tau_1)=0$:
\be
 \label{eqn:tau1}
  \tau_1  = \frac{mr}{b} \log \frac{1 + m\rho}{1 + m/4}  \,.
\ee
In this $\tau$-range the amplitudes $\xi,\eta$ oscillate and decay with damping rate $-\frac{1}{2}$.
The second regime (further exponential decrease of $\xi$ and $\eta$, but now without oscillations) ends at
$\tau_2$, defined by the condition that one of the eigenvalues~$\lambda(\tau_2)$ be zero, i.\,e.,
$\zeta(\tau_2)=0$:
\be
 \label{eqn:tau2}
  \tau_2 = \frac{mr}{b} \log (1 + m\rho)  \,.
\ee
From $\tau_2$ to $T$ one eigenvalue is positive and the variables $\xi$ and $\eta$ increase at a growing rate.
This behavior is in complete qualitative agreement with Fig.~\ref{fig:A3S4Logintime}.

\begin{figure}[H]
\centering
\includegraphics[width=0.8\textwidth]{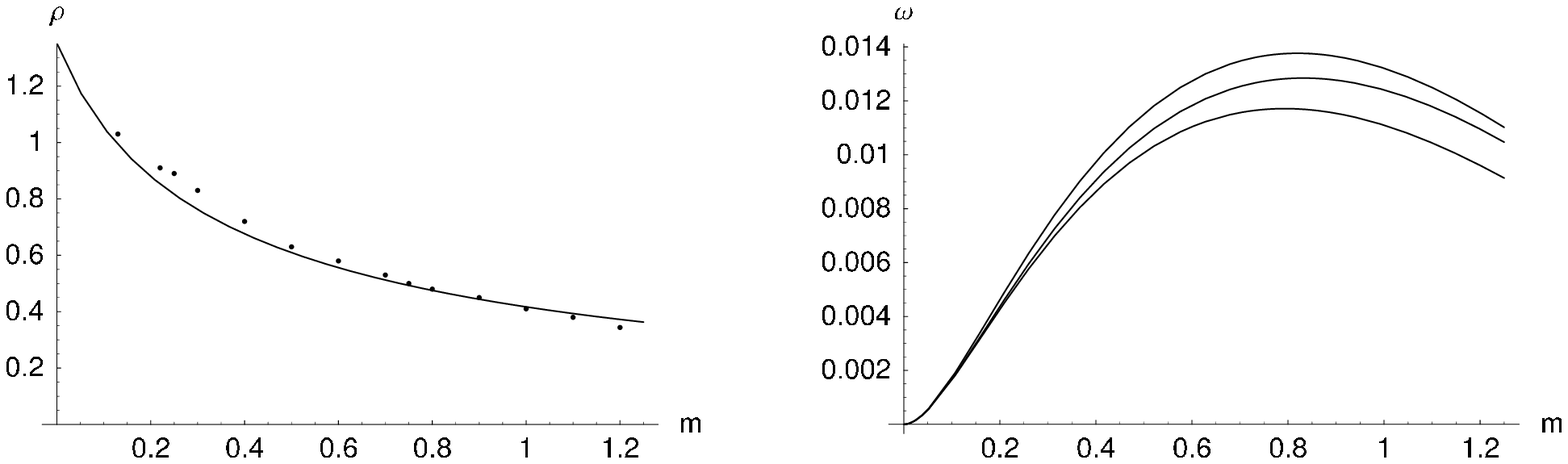}
\caption{\label{fig:Airyplots} Plots of $\text{Bi}( \alpha_0^{1/3} rt )$ (left) and
$\log|\xi(t)| = \log( \exp(-mr t/2) |\text{Bi}( \alpha_2^{1/3} rt ) | )$ (right),
where $t = \tilde \tau/r$ varies on the ($r$-independent) interval
$[- \tau_1 /r, (T - \tau_1)/r]$. The parameters are $r=1552, \sigma=1164$, corresponding to S$_4$.}
\end{figure}

With the ansatz $\xi(\tau) = e^{-\tau/2} X(\tau)$, the two linear equations of first order are transformed into
$X''  - \Delta(\tau) X  = 0$ which is, in contrast to the previous case~A, not the Airy equation because
it is not linear in~$\tau$. However, with the substitution $\tau \mapsto u$ via
$u = 2r\frac{m}{b} \exp\{-(b/2mr)\tau\}$ it may be reduced to the standard form of the Bessel equation. The
$\xi,\eta$ -solution is then a linear combination of the two Bessel functions of first and second kind of order
$-2r\frac{m}{b} \sqrt{1/4 + 1/m}$ and argument $u$. For fixed, finite $\tau$ but large $r$ we have
$u \approx 2r\frac{m}{b}-\tau$, linear in $\tau$, but with a large additive constant.
Thus in the limit of large $r$ the asymptotics for Bessel functions
with large negative order and large arguments is needed. The simplest way to obtain such asymptotics is to use
an approximation by Airy functions from the very beginning when solving the $X$-equation. This is done by
linearizing the discriminant $\Delta(\tau)$ in the original equation at an appropriate time~$\tau^\ast$,
\be
 \label{Deltalin}
     \Delta(\tau) = \Delta(\tau^\ast) + \textstyle\frac{b}{mr}\bigl(\zeta(\tau^\ast) + \frac{1}{m}\bigr) (\tau - \tau^\ast) + \ldots \, ,
\ee
upon which we obtain the Airy equation for $\tilde X (\tilde\tau) = X(\tau)$,
\be \label{eqn:X}
   \tilde{X}'' - \alpha \tilde \tau \tilde X = 0, \qquad \alpha = \textstyle\frac{b}{mr}\bigl(\zeta(\tau^\ast)+ \frac{1}{m}\bigr) \,,
\ee
where
\be \label{eq:tauast}
 \tilde\tau  = \tau - \tau_1^\ast \qquad \text{with} \qquad \tau_1^\ast = \tau^\ast - \Delta(\tau^\ast)/\alpha \, .
\ee
But what should we take for $\tau^\ast$? A natural choice seems to be $\tau^\ast = \tau_1$ where $\Delta = 0$ and $\zeta=\frac14$,
whence
\be
 \label{alpha1}
    \alpha = \alpha_1 = \frac{b}{m^2r}\Bigl( 1 + \frac{m}{4}\Bigr)  \qquad \text{and} \qquad \tau_1^\ast
= \tau_1  \qquad  \text{($\tau^\ast = \tau_1$)}.
\ee
However, if we care to treat the decaying and the exploding part of the $\xi,\eta$ -dynamics with equal weight,
then it appears natural to take the minimum of $\xi(\tau) = e^{-\tau/2} X(\tau)$ as the reference point, i.\,e.,
$\tau^\ast = \tau_2$. Here $\zeta(\tau_2)=0$ and $\Delta(\tau_2)=\frac{1}{4}$, so that from~\eqref{eqn:X} and~\eqref{eq:tauast}
we obtain
\be
 \label{alpha2}
    \alpha = \alpha_2 =\frac{b}{m^2r}  \quad \text{and} \quad \tau_1^\ast = \tau_2-\frac{m^2r}{4b}
               = \tau_1 +  O\bigl((m/4)^2\bigr) 
     \qquad \text{($\tau^\ast = \tau_2$)}.
\ee
The last step involves $\tau_2=\tau_1 + \tau_2 -\tau_1$ and an expansion
of $\tau_2-\tau_1=\frac{mr}{b}\log(1+\frac{m}{4})$ to second order in
the small quantity $\frac{m}{4}$. The difference $\tau_1^\ast -\tau_1$ is itself of second order.
Finally, when we are mainly interested in the initial phase of the decay,
$\tau^\ast = 0$ ought to be a reasonable choice. Then $\zeta(\tau^\ast)=\rho$ and
$\Delta(\tau^\ast) = -(\rho -\frac14)$ which leads to
\be
 \label{alpha0}
    \alpha = \alpha_0 = \frac{b}{m^2r}\bigl(1+ m\rho\bigr)  \quad \text{and} \quad \tau_1^\ast
            = \tau_1  + O\bigl((m/4)^2\bigr) 
    \qquad  \text{($\tau^\ast = 0$)} \ .
\ee
Again $\tau_1^\ast -\tau_1$ is of second order in small quantities.
Assuming that the $m$-corrections are not too large, we take $\tau_1^\ast = \tau_1$ in all three cases and find
the solution
\be \label{eqn:xiSol}\
  \xi(\tilde\tau) = \text{e}^{-\tau_1/2}\text{e}^{-\tilde\tau/2}\text{Bi}(\alpha^{1/3}\tilde\tau) \,,
\ee
with $-\tau_1 < \tilde \tau < T-\tau_1$. An example of this time development is given in Fig.~\ref{fig:Airyplots} where the
parameters corresponding to~S$_4$ are taken. The left picture shows Bi($\alpha^{1/3}\tilde\tau$) in the time range
$-\tau_1 < \tilde\tau \lesssim 0$, the right picture is a logarithmic plot of $\xi(\tilde\tau)$ for an entire period.
The qualitative agreement with the numerical integration of the Lorenz equations, see Fig.~\ref{fig:A3S4Logintime},
is quite impressive.

As for the value of $\alpha$, we take $\alpha_2$ to determine~$\rho$
(a glance at Fig.~\ref{fig:A3S4Logintime} shows that the gross feature of the time development is
nearly symmetric with respect to~$t_2$), and $\alpha_1$ or $\alpha_0$ for the periodicity of the oscillations in the initial phase
of the decay (which takes place between $\tau=0$ and $\tau = \tau_1$).
This is why the $\alpha$'s are different in the left and right part of Fig.~\ref{fig:Airyplots}.
The right part shows that the periodicity condition $\xi(0)=\xi(T)$, see (\ref{rhoEquation}), is approximately fulfilled for
$\text{S}_4=(r,\sigma)=(1552,1164)$ if we take $\alpha=\alpha_2$. The number of zeros, however, is not correct (6~instead of~7);
had we taken $\alpha=\alpha_0$, the values of $\ln|\xi(t)|$ would have been different between beginning and end.
On the other hand, the left part shows that with $\alpha=\alpha_0$ we get the correct number of zeros (the same is true
for $\alpha=\alpha_1$). This reflects the fact that $\Delta(\tau)$ is not a linear function of~$\tau$.

\begin{figure}
\centering
\includegraphics[width=0.8\textwidth]{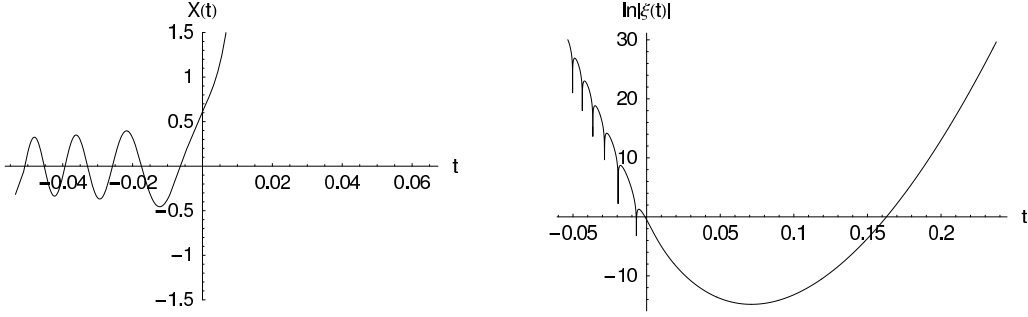}
\caption{\label{fig:rhomegavsm} Left: The amplitude $\rho(m) = \hat\rho(m)/m$ of the $\zeta$-motion as given
by~\eqref{eq:rhovsm}. The points are obtained by numerical integration of the full Lorenz equations.
Right: The frequency $\omega(m) = 2\pi/P(m)$ of the repetitive pattern in the parameter~$r$, as a function of~$m$.
From top to bottom, the curves are
calculated with $c_n = c_0$, $c_1$, and $c_2$ as defined in the text and inserted into (\ref{eq:rNfull}, \ref{eq:Pm}),
with $\rho(m)$ taken from~\eqref{eq:rhovsm} in all three cases.
}
\end{figure}

Using the asymptotics of Bi, the periodicity condition $\xi(0)=\xi(T)$ implies
\begin{equation}
 \label{expo2}
  \textstyle -\frac12 T + \frac23 \sqrt{\alpha_2} (T - \tau_1)^{3/2} = 0 \, .
\end{equation}
This leads to the following equation for the dependence of $\rho$ on $m$,
\be
  \label{eq:rhovsm}
    \sqrt{m}\log\frac{1+m\rho}{1-m\rho} = \frac{4}{3}\Bigl( \log\frac{1+m/4}{1-m\rho} \Bigr)^{3/2} \ .
\ee
The left part of Fig.~\ref{fig:rhomegavsm} is a plot of the corresponding function $\rho(m)$;
the dots are derived from numerical test values of~$\zeta$ that were generated by integrating the full Lorenz equations.
The agreement is satisfactory. In the limit of small~$m$, obtained by expanding to lowest order
in $m\rho$ and~$m/4$, we recover the result~(\ref{rhovalue}).
With increasing $m$, the amplitude~$\rho$ of the $\zeta$-motion decreases from the value $\rho_0=1.35$
to  0.493  at $m= 0.75$. For $m$ of the order of~1 the analysis becomes less and less reliable
as the linearization  of $\Delta(\tau)$, and hence the use of Airy functions, ceases to be a good approximation.

When the value $\rho(0.75) = 0.493$ is compared to the numerically determined value of $\hat\rho$, see e.\,g.\
Figs.~\ref{fig:AttractorA3} and~\ref{fig:AttractorS4}, we must remember that $\hat\rho$ was scaled with~$r$ whereas
$\rho$ is scaled with $\sigma = mr$; hence $\hat\rho = m\rho$, and $\hat\rho(0.75) = 0.37$ which agrees well with the
numerical observations.

To determine the periodicity of the $(r,\sigma)$-phase space pattern as we go along a line $\sigma=mr$
with fixed $m$ in Fig.~\ref{fig:rsigma2000},
the same arguments as in the derivation of~\eqref{wN1} lead to
\be
   \textstyle \frac23\sqrt{\alpha}\tau_1^{3/2} + \frac{\pi}{4} = N \pi \, .
\ee
But again the question arises as to which $\alpha$ should be taken. Since $\rho$ was calculated with~$\alpha_2$
it seems consistent to do the same here; the result is the lower of the three curves in the right part of
Fig.~\ref{fig:rhomegavsm}. It might be argued, however, that the behavior of the solution between $\tau=0$
and $\tau_1$ is rather determined by the slope of $\Delta(\tau)$ in that range, so we took $\alpha_1$ and
$\alpha_0$ for comparison. The three results for the $r$-values of the characteristic points A$_n$ and S$_n$
-- cf.~eq.~(\ref{Windingnumbers}) for the relation between $n$ and the winding number~$N$ --
may be summarized as follows:
\be
 \label{eq:rNfull}
   r_N(m) = \bigl(N - {\textstyle\frac14}\bigr)P(m)
\ee
with half period (in $r$)
\be
 \label{eq:Pm}
         P(m) \equiv \frac{\pi}{\omega(m)} = \frac{3\pi}{2}\frac{b}{\sqrt{mc_n}}
\Bigl(\log\frac{1+m\rho(m)}{1+m/4}\Bigr)^{-3/2} \, ,
\ee
where $c_n$ corresponds to $\alpha_n$, namely $c_0=1+m\rho(m)$, $c_1=1+m/4$, and $c_2 = 1$.
In the regime of low~$m$, all three choices of $\alpha$ give the same frequency $\omega(m)$.
With $\rho = \rho_0$ and the expansion of the $\log$~term we recover the result from the previous subsection:
$\sigma_N = mr_N = w_N\sqrt{br_N}$ with $w_N$ given by eq.~(\ref{wN}). For larger~$m$
the best agreement with the
numerical integration of the original Lorenz equations is obtained from~$c_0=1+m\rho(m)$: with the
choice $m=0.75$ the prediction for the period of the pattern
is ~$2P(m_0) = 459.7$. This is what was used to define the points S$_n$ and A$_n$
in Fig.~\ref{fig:rsigma2000}, namely, $r_N = 230 (N-\frac14)$.

\begin{figure}
\centering
\includegraphics[width=0.6\textwidth]{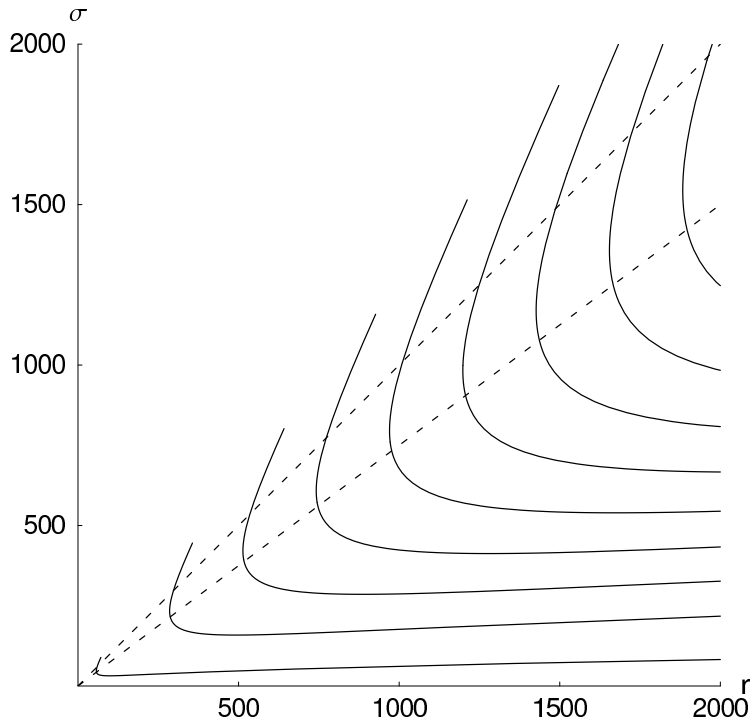} 
\caption{\label{fig:PatternTest} The curves $(r_N,\sigma_N)=\bigl( (N-\frac34)P(m), mr_N\bigr)$ for
$0<m<1.25$ and $N=1 ... 9$. The dashed lines are $\sigma = 0.75\, r$ and $\sigma = r$.
}
\end{figure}

If instead of $N - \frac14$ we take $N - \frac34$ in~(\ref{eq:rNfull}),
we obtain values $r_N(m)$ corresponding to the lower border of the $r$-bands. These are
plotted in Fig.~\ref{fig:PatternTest},
for $N$ from 1 to~9 and parameters~$m$ ranging from 0 to approximately~1.25. The agreement with
Fig.~\ref{fig:rsigma2000} is qualitatively convincing and quantitatively satisfactory.

\section{Bifurcations and coexistence of attractors}
\label{secBifs}

While the periodic orbits discussed in the previous two sections form the backbone of the structure in the
$(r,\sigma)$-phase diagram on large scales of $r$ and~$\sigma$, they do not represent the whole picture.
In this section we report on the scheme of bifurcations that connects the different bands, and we also
give a short discussion of the coexistence of attractors.

\subsection{Types of attractors: periods and symmetry}
\label{ssec:Types}

In addition to the periodic attractors of periods~1 and~2, there are small bands of stability for
periodic orbits of all periods, and of chaotic motion. This is summarized, using a color code, in
Figs.~\ref{fig:r409sigma60} and~\ref{fig:rsigma2000}.
Within a given band, the attractors are the same type, i.\,e., they share the same period
and symmetry property with respect to the reflection~$\Sigma$, see~(\ref{symmetry1}).
We use the notation ``ps'' for p-periodic symmetric attractors, and ``pa'' for  p-periodic
asymmetric attractors (whose mirror images under~$\Sigma$ are also attractors).

\begin{figure}[h]
\centering
\includegraphics[width=\textwidth]{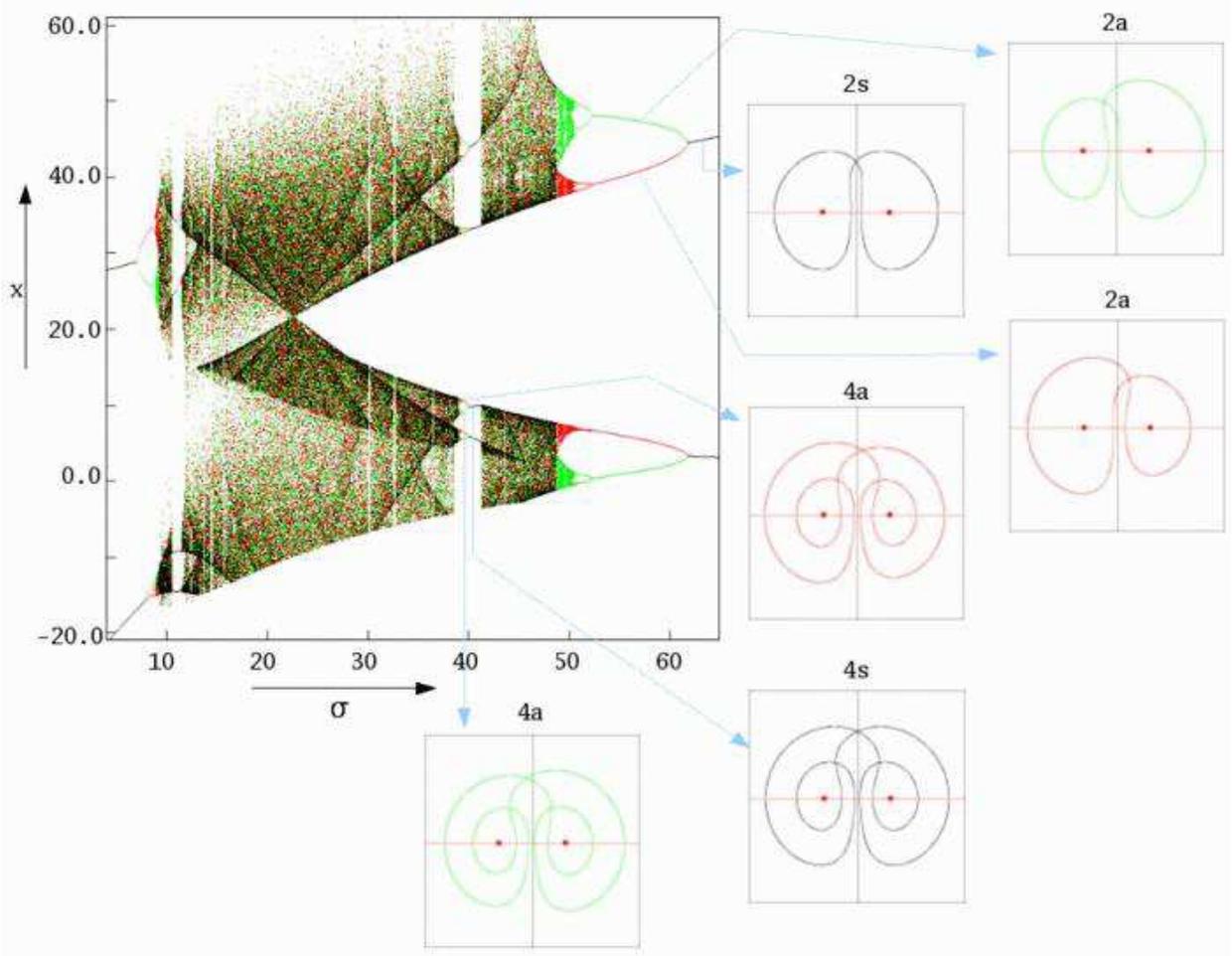}
\caption{\label{sigmaxbifdia} $(\sigma,x)$-bifurcation diagram at fixed $r = 178$, $b =8/3$ for $4 < \sigma < 65$.
The initial conditions are $x_0 = y_0 = 0.001, z_0 = r - 1$ (red) and $x_0 = y_0 = -0.001,
z_0 = r - 1$ (green). Attractors are asymmetric when they carry only one of the colors red or green. There are
also symmetric attractors, given in black (superposition of red and green, as they are reached from both
initial conditions). A few periodic orbits are shown in $(y,z)$-projection, of periods~2 in the upper right part,
and periods~4 in the lower right; arrows indicate the corresponding branches of the bifurcation diagram.
The conspicuous merging of the two major chaotic bands, as $\sigma$ is lowered, seems to occur at or very close to
the value $\sigma_s(r)  = \sqrt{br} = 21.8$.
}
\end{figure}

As the structural features of the $(r,\sigma)$-phase diagram are different in the regions
of low and high~$\sigma$ (separated roughly by $\sigma_s(r)  = \sqrt{br}$),
we discuss them in turn. To begin with, we present in Fig.~\ref{sigmaxbifdia}
a survey on the bifurcation behavior of attractors along the line $r = 178$ for
$4 < \sigma < 65$ .
This $\sigma$-range bridges the two regions of small and large Prandtl numbers, respectively. It starts well below
$\sqrt{br}$ ($= 21.8$ for this~$r$) and ends far above.
The bifurcation scheme along this line is shown in the left part of Fig.~\ref{sigmaxbifdia}.
It records the $x$-values of the attractors in the Poincar\'e surface of section
$z=r-1$. Red points were computed with initial conditions $x_0 = y_0 = 0.001, z_0 = r - 1$, green points with
$x_0 = y_0 = -0.001, z_0 = r - 1$. When the colors do not mix, they represent asymmetric attractors; symmetric attractors
are reached from both initial conditions and appear in color black.
Starting with the highest~$\sigma$, we observe a small black segment of symmetric orbits with period~2,
i.\,e., the attractors are type~2s. The $(y,z)$-projection for $\sigma = 63.5$  is shown at the right side.
Lowering $\sigma$ to approximately~62, the 2s-orbit
splits into two 2a-orbits, one red, the other green, mirror images of each other under~$\Sigma$. This is illustrated
for $\sigma = 57$. Each of these two attractors undergoes a period doubling scenario,
2a $\rightarrow$ 4a $\rightarrow$ 8a $\rightarrow$ \ldots, corresponding to red, green, blue bands in
Fig.~\ref{fig:r409sigma60}, until chaos appears at about $\sigma = 50.5$.
When $\sigma$ is further decreased, the inverse cascade of band merging as described in~\cite{Gr77} is observed.
It ends at $\sigma \approx 48.5$, where the two bands of asymmetric chaos merge into one band of symmetric chaos, thereby
restoring the symmetry under~$\Sigma$.

As $\sigma$ falls below approximately~41.2, a period~4 window emerges by an intermittency scenario. At first the attractor
is symmetric, of type~4s, as shown for $\sigma = 40.7$. Again, this orbit undergoes a pitchfork
bifurcation into two asymmetric orbits of period~4, shown in red and green for $\sigma = 39.7$, and then the
period doubling sequence takes over for both of them. In Fig.~\ref{fig:r409sigma60}, this window
appears as the dominant green band in the white chaotic region.

\begin{figure}[h]
\centering
\includegraphics[width=0.9\textwidth]{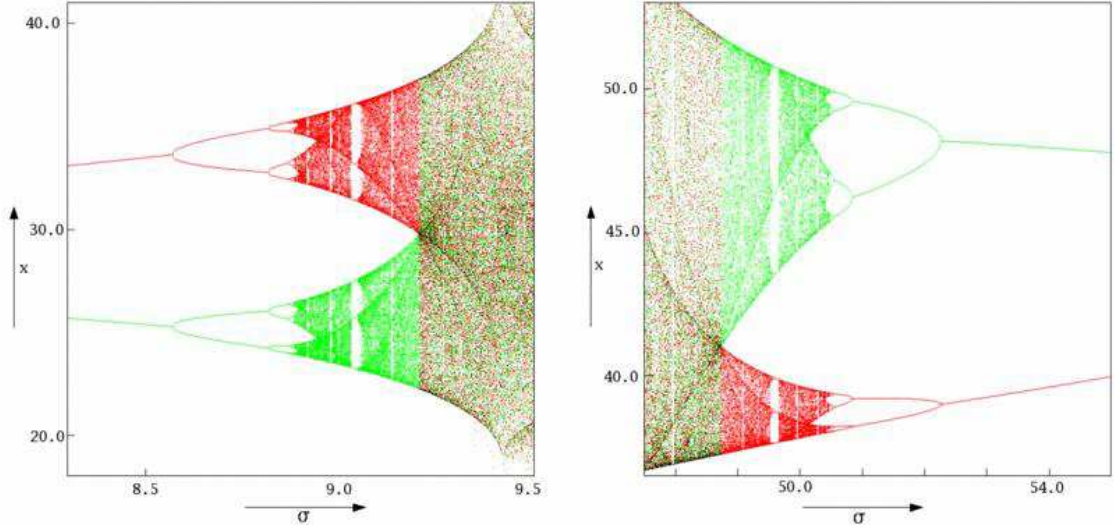}
\caption{\label{fig:Blowups} Blowups of two pieces of the upper part of Fig.~\ref{sigmaxbifdia}. Left: $8 < \sigma < 9.5$,
upper left branch;
right: $45 < \sigma < 55$, upper right branch. The two pictures are like mirror images of each other. Again, red and green
indicate the asymmetric attractors, approached from the two $\Sigma$-images of the initial values.
}
\end{figure}

\begin{figure}[h]
\centering
\includegraphics[width=0.45\textwidth]{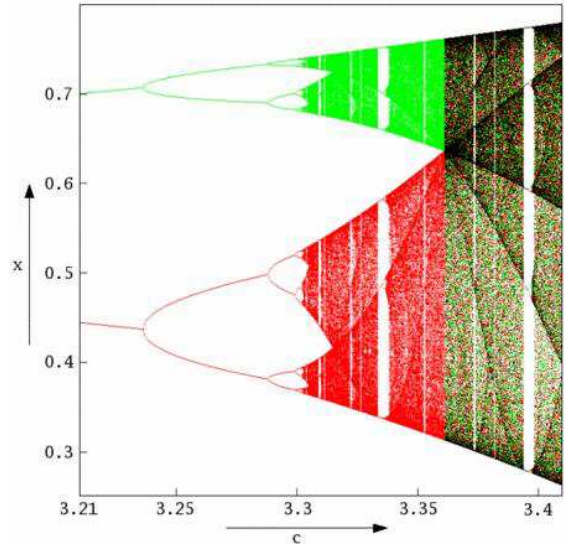}
\caption{\label{fig:cubic} Detail of the $(c,x)$-bifurcation diagram of the cubic iteration~\eqref{eq:cubic}.
}
\end{figure}

Fig.~\ref{fig:Blowups} shows the upper part of the bifurcation diagram in Fig.~\ref{sigmaxbifdia} in greater detail,
for the intervals $45 < \sigma < 55$ and
$8.0 < \sigma < 9.5$. Except for their different scales
the pictures are much like mirror images of each other. In both cases the scenarios coincide in many aspects
(though not completely) with the bifurcation scheme of the antisymmetric cubic map
\be
x_{n+1} = f(x_n) = (1-c)x_n + cx_n^3  \  , \qquad -1 \le x \le 1 \ ,
\label{eq:cubic}
\ee
shown in Fig.~\ref{fig:cubic} in the ranges $3.2 \lesssim c \lesssim 3.4$ and $0.25 \lesssim x \lesssim 0.8$.
This similarity was pointed out and explored using symbolic dynamics
in~\cite{Di90} and~\cite{Fa96}. The reason for the good correspondence seems to be that~\eqref{eq:cubic} is the
simplest polynomial 1-dim map that shares with~(\ref{lorenz}) a reflection symmetry. The invariance with
respect to $\Sigma$ is here represented by the symmetry under $x \mapsto \tilde\Sigma x =-x$, namely
$\tilde\Sigma^{-1} f \tilde\Sigma = f $. This modifies the standard period doubling scenario of
the quadratic map $x \mapsto ax(1-x)$ in such a way that the first bifurcation of a symmetric orbit
of period~p is not a period doubling ps $\rightarrow$ (2p)s but rather a pitchfork bifurcation where ps
breaks up into pa and its mirror image $\tilde\Sigma\text{(pa)}$. Only then does
period doubling start: pa $\rightarrow$ (2p)a $\rightarrow$ (4p)a $\rightarrow$ \ldots\,.

The character of the bifurcation diagram Fig.~\ref{sigmaxbifdia} stays the same as we increase~$r$
and take any $\sigma$-segment that connects the lowest two 2s-bands across the line $\sigma_s(r) = \sqrt{br}$.

\begin{figure}[h]
\centering
\includegraphics[width=\textwidth]{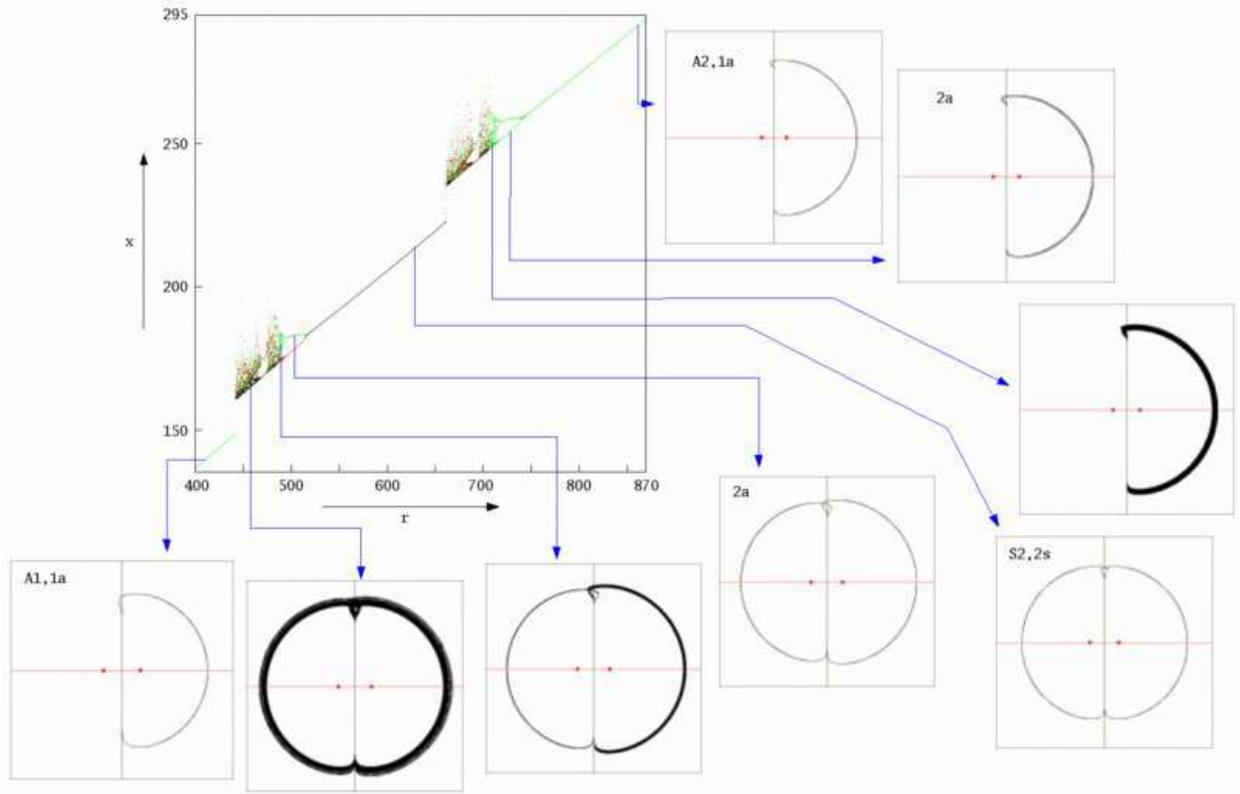}
\caption{\label{fig:rxbifdia} Upper left: $(r,x)$-bifurcation diagram with $m=\sigma/r = 0.75$ and
$r = 400 \ldots 870$, $b =8/3$.
The initial conditions are $z_0=r+1000$ and $x_0 = y_0 = 10^{-5}$ (red) or $x_0 = y_0 = -10^{-5}$ (green).
The 8~small windows show attractors corresponding to the parameter values identified by the arrows, in
$(y,z)$-projection. Following the sequence in clockwise direction, starting at the top, the bifurcation scheme between A$_2$
($r = 863$) and A$_1$ ($r = 403$) is illustrated: 1a-orbits (yellow band with center A$_2$, see Fig.~\ref{fig:rsigma2000})
undergo period doubling, then become chaotic. Out of this chaos, 2s-orbits emerge (red band with center S$_2$) and undergo a
pitchfork bifurcation into 2a and its $\Sigma$-image; these bifurcate according to the period doubling scenario before
chaos sets in, first asymmetric, then symmetric. Finally, 1a-orbits (of the A$_1$-band) emerge and the scenario starts anew.}
\end{figure}

Let us now consider what kind of regularity we
can detect in the region of large $r$ and~$\sigma$, in the onion like structure of Fig.~\ref{fig:rsigma2000}.
A representative picture of the bifurcation behavior is obtained by considering any parameter interval
that connects two consecutive bands of equal symmetry. We take the line from A$_1$ to A$_2$ and show
the corresponding $(r,x)$-bifurcation diagram in the upper left part of Fig.~\ref{fig:rxbifdia},
with the two initial conditions
$x_0 =y_0 = 10^{-5}$ (red) and $x_0 =y_0 = -10^{-5}$ (green); $z_0 = r+1000$ in both cases.
(This large value of $z_0$ is to a large extent arbitrary; however, it was observed that chaos coexists with
periodic orbits, and that initial conditions with $z_0$ of order $r$ or lower tend to be captured by the
latter.) The Poincar\'e surface of section is $z_0 = r -1$ as usual.
Only the branch with $x>0$ is shown: the complete picture contains the mirror image $x \mapsto -x$,
with colors red and blue interchanged. The small windows in Fig.~\ref{fig:rxbifdia} present a number of
characteristic attractors in $(y,z)$-projection, arrows connecting to the respective points on the
bifurcation diagram. On the upper branch starting from~A$_2$, we have asymmetric periodic attractors of type~1a.
These undergo a period doubling scenario (orbit 2a, etc.), including the inverse cascade (with asymmetric
chaotic orbits), before they
merge with their image under~$\Sigma$ and yield two symmetric chaotic orbits. This scenario is illustrated with the
orbits in the upper right part of the picture.
Upon lowering~$r$, the chaotic attractor suddenly
disappears, and an orbit of type~2s emerges, characteristic of the red band with center~S$_2$. This symmetric orbit splits into
two asymmetric orbits of type~2a, in a pitchfork bifurcation like in Fig.~\ref{sigmaxbifdia}
(whose rightmost part belongs to the band~S$_1$ rather than~S$_2$). Next, period doubling sets in again.
The chaos in the range of inverse cascading is at first asymmetric (though different from the one in the yellow band),
but turns into symmetric chaos at the end point of the band merging. Finally, an attractor of type~1a emerges
from this chaos and gives rise to a picture around A$_1$ similar to the one with which we started at~A$_2$.

We have not investigated the mechanism by which the ``new'' stable periodic orbits are generated when
the chaotic motion becomes unstable
upon lowering~$r$. In fact, they coexist with the chaotic attractors before they are seen in Fig.~\ref{fig:rsigma2000}.
We cannot tell whether they emerge in an intermittency scenario. It may be that their range of existence,
ignoring the question of their stability, extends all the way up to infinity.

\begin{figure}[H]
\centering
\includegraphics[width=0.8\textwidth]{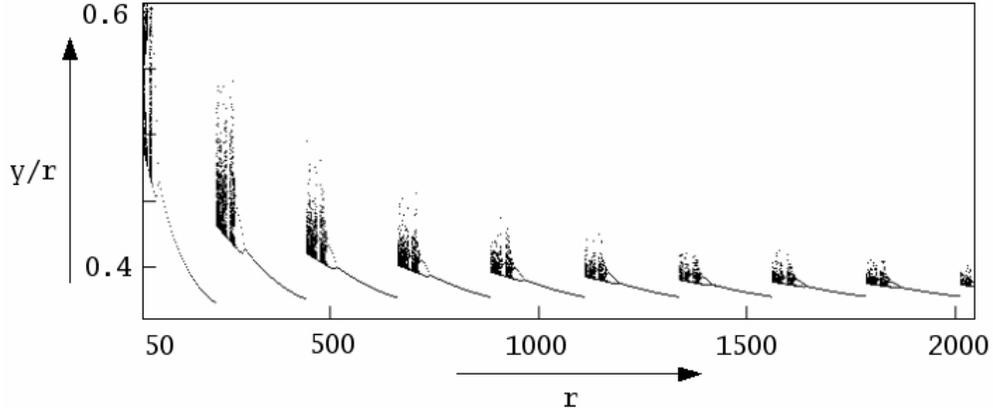}
\caption{\label{fig:rhocheck} $(r,y/r)$-bifurcation diagram with $m=\sigma/r = 0.75$ and
$50 < r < 2050$. The picture shows that the repetitive structure of Fig.~\ref{fig:rsigma2000} includes
the bifurcation schemes.
}
\end{figure}

Fig.~\ref{fig:rhocheck} is an extended version of the bifurcation diagram in Fig.~\ref{fig:rxbifdia}.
The range of $r$ is from 50 to 2050, and instead of the ordinate~$x$ we plot~$y/r$, again for $z=r-1$.
Remember that for a given periodic orbit, $y$ reaches its maximum value when $z$ is near~$r$.

The picture contains two messages: (i) the repetitive nature of the pattern in
Fig.~\ref{fig:rsigma2000} extends to the bifurcation scenarios, and (ii) the value of $y/r$ at its maximum,
for periodic attractors, seems to tend to a constant $\hat\rho = \rho(m)/m$
as $m=\text{const}$ and $r\rightarrow\infty$.
This was an essential assumption in the scaling scheme that we used in the analysis of Section~\ref{ssecm}.
The value $\hat\rho = 0.37$ that our theoretical investigation gave for $m=0.75$ is close to the observed limit.


\subsection{Coexistence of attractors}
\label{ssec:Coexistence}

\begin{figure}[H]
\centering
\includegraphics[width=\textwidth]{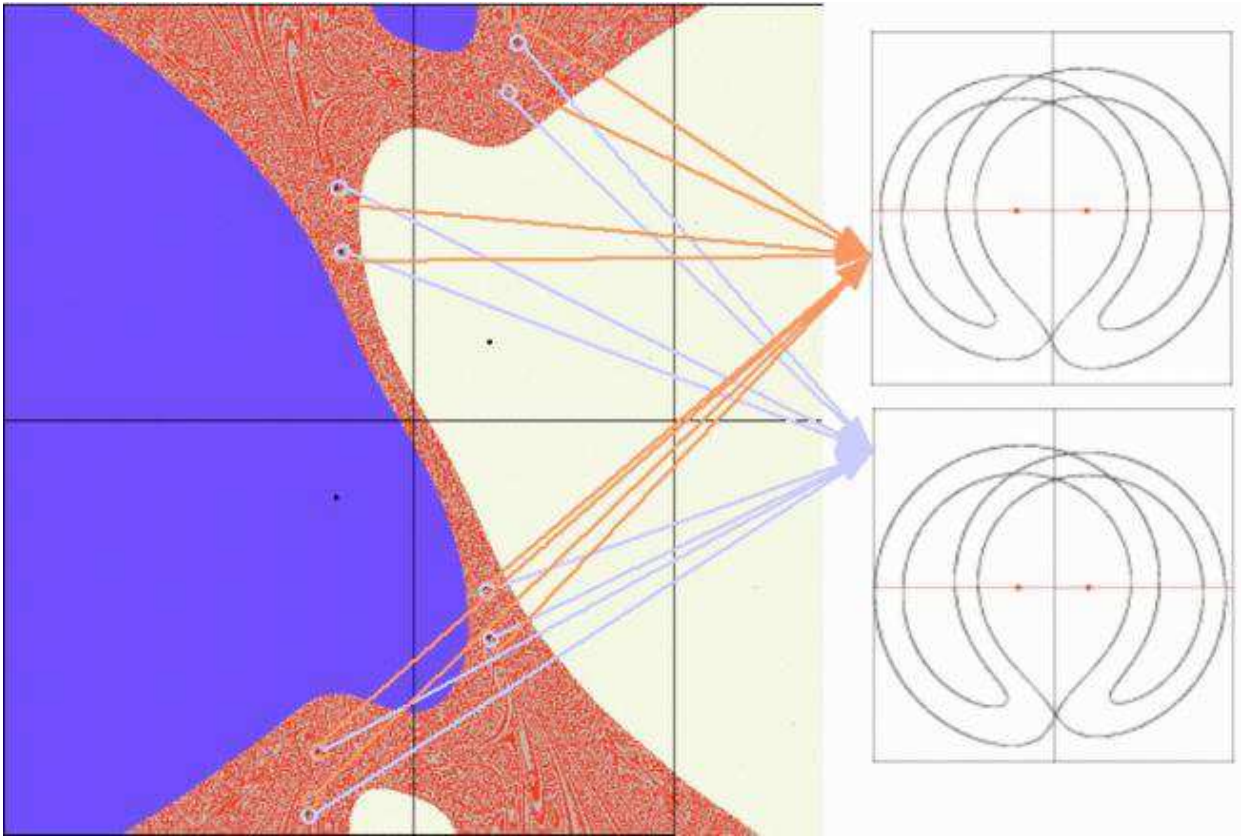}
\caption{\label{fig:4basins}
Left: Basins of attraction of 4 coexisting attractors, for $(r,\sigma) = (108.82, 3.6)$.
Initial conditions in the $(x_0,y_0)$-plane, $-91 < x_0,y_0 < 91$,
with $z_0 = r-1$, are scanned pixel by pixel and given a color depending
on the attractor to which they are asymptotic: white for P$_+$, blue for P$_-$, red or gray for the
asymmetric period~4 orbits shown on the right (abscissa $y$ and ordinate $z$). The fixed point attractors P$_\pm$ are
marked as black dots.
All intersection points (8 each) of the two period~4 attractors with the Poincar\'e surface $z = r-1$ are emphasized.
}
\end{figure}

It has already been mentioned several times that different attractors may coexist. For example, whenever
an attractor~A is not in itself symmetric under~$\Sigma$, it coexists with its partner~$\Sigma$A.
This applies in particular to the fixed point attractors P$_\pm$ in the black region, or to the
1a-attractors in the yellow bands. Consequently, there must be at least four coexisting attractors where
the yellow bands extend into the black region, as can be seen in Fig.~\ref{fig:rsigma2000}.
A less trivial case is analyzed in Fig.~\ref{fig:4basins}.
The parameters  $(r,\sigma) = (108.82, 3.6)$ are located in the lower part of Fig.~\ref{fig:r409sigma60} where
the green asymmetric period~4 band overlaps with the black region.
The left part of Fig.~\ref{fig:4basins} shows the Poincar\'e surface of section $z=r-1$,
with initial conditions $(x_0,y_0)$ colored pixel by pixel according to the final state to which they
are attracted: white or blue if the final state is the fixed point P$_\pm = (\pm 16.96,\pm 16.96,107.82)$, respectively,
red or gray if it is one of the two 4a-orbits.
No other attractor besides these four was found.
The blue and white basins of the point attractors P$_\pm$ appear to have smooth boundaries
whereas the boundary between the basins of attraction of the two
4a-orbits is highly entangled.

\begin{figure}[H]
\centering
\includegraphics[width=\textwidth]{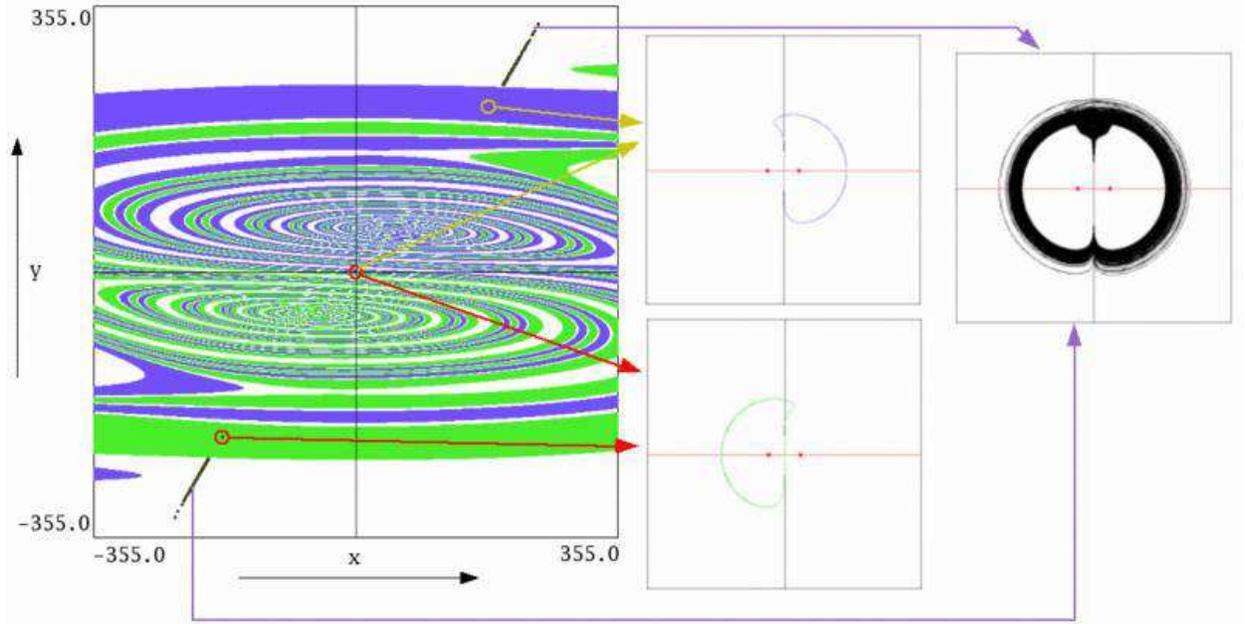}
\caption{\label{fig:3basins}Left: Basins of attraction of three coexisting attractors for $(r,\sigma) = (1200, 255.64)$.
Initial conditions with $-355 < x_0, y_0 < 355$, $z_0 = r-1$ are scanned and colored
according to their final fate.
The three attractors are shown on the right, in $(y,z)$-projection:
two asymmetric orbits of period~1 (with the blue and green basins of attraction),
and one symmetric chaotic attractor (in black, with the white basin of attraction). The intersections of the attractors
with the Poincar\'e surface are emphasized by arrows.
}
\end{figure}

Another case of coexisting attractors is documented in Fig.~\ref{fig:3basins} where
the parameters $(r,\sigma) = (1200, 255.64)$ are from the chaotic band at the upper rim
of the lowest yellow band in Fig.~\ref{fig:rsigma2000}.
Fig.~\ref{fig:3basins}
demonstrates how a symmetric chaotic attractor coexists with two attractive asymmetric period~1 orbits.
No other attractors have been detected.
The basin boundaries seem to be complicated but not fractal; they are spiralling out from the unstable fixed points~P$_\pm$.
Each point is colored according to which attractor absorbs it.
Each basin trivially contains its respective attractor.
The location of their intersections with the plane $z=r-1$ is marked by black dots.

\begin{figure}[H]
\centering
\includegraphics[width=\textwidth]{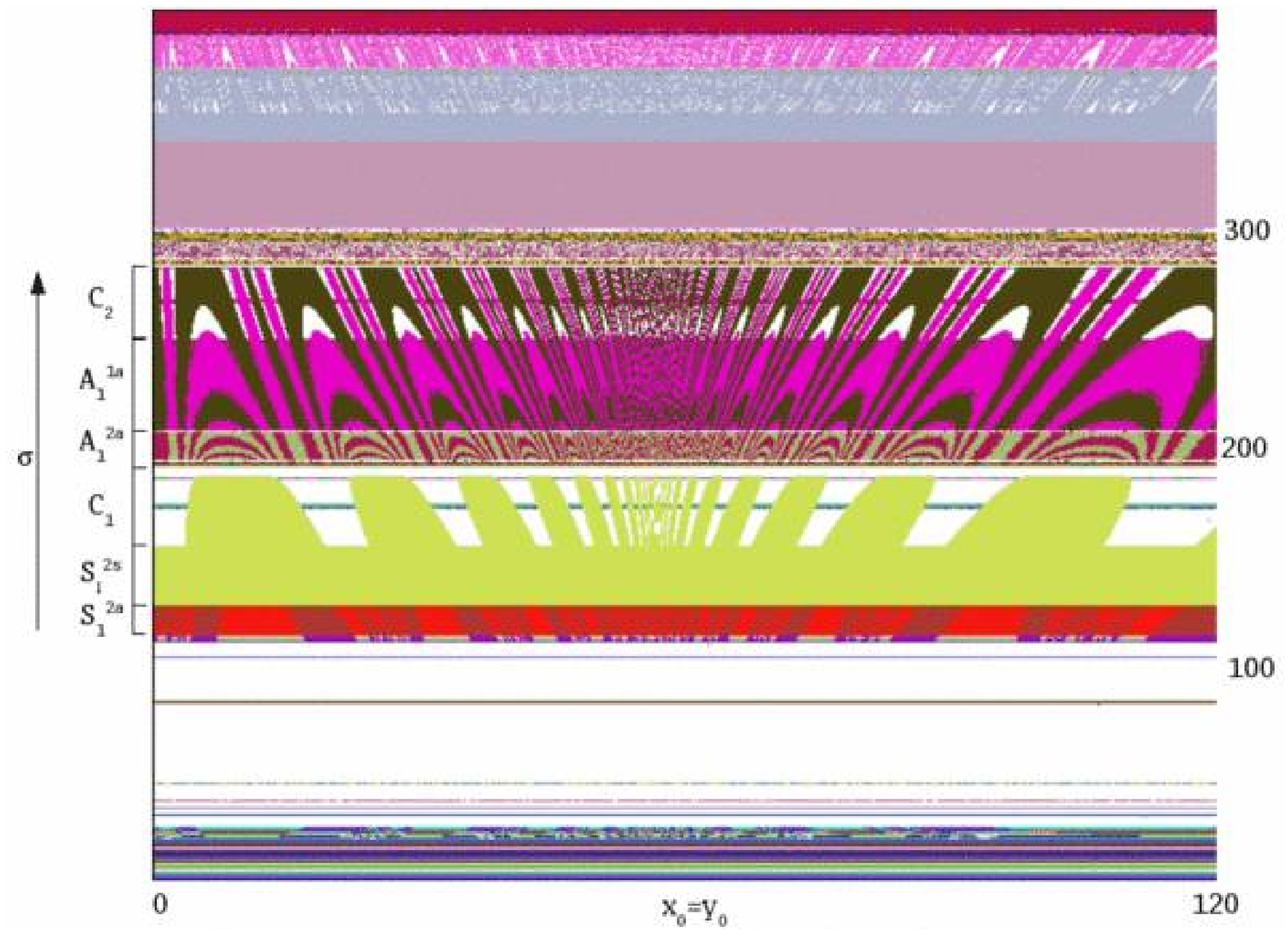}
\caption{\label{fig:AttractorCoexistence} $\sigma$-stack of domains of attraction, for $r=1200$. For $\sigma$
in the range between $0$ and $400$ (ordinate) initial points along the diagonal $x_0=y_0$ (and in the Poincar\'e plane
$z=r-1$) are given a color according to which orbit attracts them. White, in particular, means there are chaotic orbits.
The values $x_0=y_0$ are taken from the range $(0,400)$.
The features of the $\sigma$-ranges S$_1^{2a}$, S$_1^{2s}$, C$_1$, etc. are explained in the text.
}
\end{figure}

Inspection of Figs.~\ref{fig:4basins} and \ref{fig:3basins} suggests that a scan
of initial conditions along the line $x_0 = y_0$, $z_0=r-1$,
might catch all (co-)existing attractors. Thus instead of scanning a two-dimensional
set of initial conditions, we scan along this diagonal line only and use the second dimension
to superimpose these scans for a range of parameters~$\sigma$.
Such a ``$\sigma$-stack'' is shown in Fig.~\ref{fig:AttractorCoexistence} for $r=1200$. The abscissa is
$x_0 = y_0$ in the range from $0$ to $120$, and the ordinate is $\sigma$ from $0$ to $400$.
Comparison with Fig.~\ref{fig:rsigma2000} shows that this $\sigma$-range covers the bands containing
the points S$_1$, A$_1$, and S$_2$, in addition to the low-$\sigma$ structure with regular behavior at
the very bottom and chaos in a range around $\sqrt{br} = 56.6$.

The coloring scheme is somewhat arbitrary
and was designed so that the coloring of different attractors could be done
automatically. For a fixed value of $\sigma$
different color designates different attractors, i.\,e., the number of different colors
on a scan-line is the number of coexisting attractors. Same color for adjacent points
with different~$\sigma$ implies that the corresponding attractors are deformations
of each other. Color white is reserved for chaotic motion.

Let us concentrate on the $\sigma$-ranges which are marked as S$_1^{2a}$, S$_1^{2s}$, C$_1$, etc.
S$_1^{2s}$ is the $\sigma$-interval $(126.5, 154.8)$. The attractors are symmetric orbits of period~2 and
type S$_1$, see Fig.~\ref{fig:Winding}, and there are no other attractors; hence there is only one color.
For smaller $\sigma$ in the range S$_1^{2a}$, $113 < \sigma < 126.5$, the attractors have undergone a pitchfork bifurcation
and exist in pairs of asymmetric orbits of period~2. Thus we see two colors, corresponding to the two
basins of attraction. At larger $\sigma$, on the other hand, in the range $\text{C}_1 = (154.9,190.3)$, the attractors
of type~2s coexist with chaotic attractors; again we observe two colors.

The band with orbits of type A$_1$ exists in the range $\text{A}_1^{1a} = (207.3, 250.8)$.
As the attractors are asymmetric orbits, they occur in pairs, and two colors appear (black and magenta).
At lower $\sigma$, we have a range $\text{A}_1^{2a} = (190.3, 207.3)$ where period doubling has increased the period
but not the number of attractors; thus there are again two colors. At larger $\sigma$, in the range
$\text{C}_2 = (250.8, 283)$, the two asymmetric periodic attractors coexist with a chaotic attractor, and three colors
are needed. Fig.~\ref{fig:3basins} is an example from this range.

If $\sigma$ is further increased, we enter the regime which is discussed in Fig.~\ref{fig:rxbifdia}. The bifurcation tree of
the S$_2$-band leaves a complicated trace in Fig.~\ref{fig:AttractorCoexistence}, but then there comes a range where a 2s-orbit
is the system's only attractor. From thereon, the pattern seems to repeat itself, but the details on the $x=y$ axis become
more and more intricate.

\clearpage



\end{document}